\documentclass[fleqn,usenatbib]{mnras}

\usepackage{newtxtext,newtxmath}

\usepackage[T1]{fontenc}

\DeclareRobustCommand{\VAN}[3]{#2}
\let\VANthebibliography\thebibliography
\def\thebibliography{\DeclareRobustCommand{\VAN}[3]{##3}\VANthebibliography}

\usepackage{graphicx}	
\usepackage{amsmath}	
\usepackage{enumitem}
\usepackage{ctable} 
\usepackage{soul}

\newcommand{\hi}{\text{H\textsc{i}}}

\newcommand{\zeq}{z_\text{eq}}
\newcommand{\secref}[1]{\hyperref[#1]{Section~\ref*{#1}}}
\newcommand{\appref}[1]{\hyperref[#1]{Appendix~\ref*{#1}}}

\title[Detecting the power spectrum turnover with \hi\ IM]{Detecting the power spectrum turnover with \hi\ intensity mapping}

\author[S. Cunnington]{Steven Cunnington$^{1,2}$\thanks{E-mail: steven.cunnington@ed.ac.uk}
\\
$^{1}$Institute for Astronomy, The University of Edinburgh, Royal Observatory, Edinburgh EH9 3HJ, UK\\
$^{2}$School of Physics \& Astronomy, Queen Mary University of London, 327 Mile End Road, London E1 4NS, UK\\
}

\date{Accepted XXX. Received YYY; in original form ZZZ}

\pubyear{2020}

\begin{document}
\label{firstpage}
\pagerange{\pageref{firstpage}--\pageref{lastpage}}
\maketitle

\begin{abstract}
A goal for pathfinder intensity mapping (IM) surveys will be detecting features in the neutral hydrogen (\hi) power spectrum, which serve as \textcolor{black}{conclusive evidence} of cosmological signals. Observing such features at the expected scales in \hi\ IM auto-correlations, where contribution from systematics is uncertain, will provide a more convincing cosmological detection. We demonstrate how the turnover, i.e. the peak of the power spectrum at ultra-large scales, can be detected with \hi\ IM. We find that a MeerKAT 4,000$\,\text{deg}^2$ survey using the UHF-band is capable of a $3.1\sigma$ detection of the turnover, relative to a null model power spectrum with no turnover. This should exceed what is capable by current galaxy surveys in optical and near-infrared. The detection significance falls to ${\sim}1\sigma$ in MeerKAT's L-band but can reach ${\sim}13\sigma$ with the SKAO, which should easily surpass the constraints from future Stage-IV-like spectroscopic galaxy surveys. We also propose a new model-independent methodology for constraining the precise turnover scale ($k_0$) and our tests on UHF-band simulated data achieved a precision of 10\%. This improved to 2.4\% when using the full SKAO. We demonstrate how the results are robust to foreground contamination by using transfer functions, even when an incorrect cosmology has been assumed in their construction. Given that the turnover is related to the horizon scale at matter-radiation equality, a sufficiently precise constraint of $k_0$ presents the possibility for a novel probe of cosmology. We therefore present a potential methodology for constructing a standard-ruler-based distance measurement, independent of the sound horizon, using the turnover location in the \hi\ power spectrum.

\end{abstract}




\section{Introduction}

The standard cosmological model ($\Lambda$CDM) has the majority of its parameters constrained to within sub-percent levels \citep{Aghanim:2018eyx}. Few can therefore deny that we are in an era of precise cosmological analysis. However, there still exist statistically significant tensions between different cosmological probes \citep{Verde:2019ivm, Knox:2019rjx, Joudaki:2019pmv}. A way to potentially shed light on these tensions is through novel ways of measuring the same parameters with different techniques to either rule out or upweight certain theoretical explanations.

A novel way of probing cosmic large scale structure is by mapping the unresolved diffuse redshifted 21cm signal from neutral hydrogen (\hi) using radio telescopes, a process known as \hi\ intensity mapping (IM) \citep{Bharadwaj:2000av,Battye:2004re,Wyithe:2007rq,Chang:2007xk}. The set of systematics associated with \hi\ IM will be vastly different from galaxy surveys with optical and near-infrared telescopes, conventionally used to probe large scale structure. Thus, if similar cosmological tensions are concluded with \hi\ IM, then more confidence can be placed on the assumption that such conclusions have not been influenced by systematics. It has been previously highlighted that radio surveys should be particularly adept at probing ultra-large cosmological scales, most efficiently performed using the \hi\ IM technique \citep{Camera:2013kpa,Bull:2014rha,Alonso:2015uua,Shi:2015tje,Smoot:2014oia,Kovetz:2017agg}. This opens up potential new ways to probe cosmology from the largest scales.

Planned wide and deep \hi\ IM surveys with e.g. the Square Kilometre Array Observatory (SKAO)\footnote{\href{https://www.skatelescope.org/}{skatelescope.org}} \citep{Bacon:2018dui} will efficiently map large cosmic volumes without the disadvantages of high shot-noise or poor redshift calibration, as experienced from spectroscopic or photometric galaxy surveys respectively. Furthermore, line-intensity mapping surveys in general should be more efficient for surveying high redshifts \citep{Bernal:2019gfq} compared to their optical counterparts. Pathfinder observatories are already conducting IM observations \citep[e.g. MeerKAT, SKAO's precursor][]{Wang:2020lkn} and wide-area surveys \citep[e.g. MeerKLASS][]{Santos:2017qgq} will soon be performed which will resolve the size of modes required to probe cosmological information on the largest scales.

Whilst detection of cosmological power spectra have been made with \hi\ IM \citep{Masui:2012zc,Wolz:2015lwa,Anderson:2017ert,Wolz:2021ofa}, these have all relied on cross-correlations with overlapping galaxy surveys and have only been able to constrain the effective amplitude of the power spectrum on a small range of scales. In auto-correlation the contribution from residual systematics to the amplitude is less certain, thus early pathfinder surveys could struggle to determine when systematics have been reduced enough to make a detection. Distinctive features in the power spectrum at the expected location represent \textcolor{black}{conclusive evidence} of cosmological signal, thus observing them provides a robust detection, less likely influenced by systematics. A near-term aim for \hi\ IM pathfinder surveys will therefore be to detect known cosmological features in the power spectrum. 

The most studied feature is a series of wiggles in the power spectrum caused by baryon acoustic oscillations (BAO) in the early-Universe's photon-baryon fluid which at recombination imprint a preferred scale of matter clustering \citep{Percival:2001hw,Blake:2003rh}. However, cosmological surveys with SKAO and its pathfinders will rely on single-dish intensity mapping \citep{Battye:2012tg} to probe the largest scales. This means the density field will be mapped with low angular resolution ($\gtrsim 1\,\text{deg}$) due to the large beam associated with the relatively small $15\,\text{m}$ SKAO dishes. This poor angular resolution has been shown to pose challenges for SKAO-related \hi\ IM experiments aiming to detect BAO \citep{Villaescusa-Navarro:2016kbz,Kennedy:2021srz,Avila:2021wih,Rubiola:2021afc}. A power spectrum feature on larger scales than the BAO would be better suited for detection by \hi\ IM if a sufficiently wide survey is commissioned, as planned with MeerKLASS \citep{Santos:2017qgq}. 

Fortunately, the power spectrum does contain a second distinctive feature; a broad maximum peak in power amplitude at an approximate wavenumber of $k=0.016\,h/\text{Mpc}$. We refer to this second, less studied feature as the \textit{turnover} which provides the opportunity to probe the epoch of matter-radiation equality. The present-day matter power spectrum evolved from the primordial power spectrum proportional to $k^{n_\text{s}}$ where $n_\text{s}{\sim} 0.97$ \citep{Aghanim:2018eyx}. 
Initially, the Universe is radiation dominated meaning baryons, which are coupled to photons, do not cluster due to radiation pressure and thus perturbations in cold dark matter grow at a slow logarithmic rate. Hence, structure growth is impeded during the epoch of radiation domination. As the Universe evolves and reaches the epoch of matter-radiation equality, the suppression of small scale perturbations ceased. The smaller a particular mode is, the earlier it entered the horizon, spending more time in a radiation dominated era, experiencing more retarded growth. Hence, the power spectrum is a decreasing function of $k$ on small scales with the turnover forming at a scale corresponding to the horizon size at matter-radiation equality. A higher abundance of matter will change the point at which matter-radiation equality occurs and so the turnover feature is sensitive to the matter density $\Omega_\text{m} h^2$, along with other parameters, making it a viable probe of cosmology. We refer the reader to \citet{Eisenstein:1997ik,Dodelson2003Book} for more detail.

The turnover feature has been probed in galaxy redshift surveys for decades \citep{BaughEfstathiou1993,BaughEfstathiou1994} and the information contained therein has been utilised to sharpen cosmological constraints in full-shape power spectra analyses \citep{Tegmark:2006az,Reid:2009xm}. Furthermore, using the precise turnover scale location, $k_0$, as a direct probe, has been investigated with simulations \citep{Blake:2004tr,Prada:2011uz,Pryer:2021cut} and has also been measured in WiggleZ Dark Energy Survey data \citep{Drinkwater:2009sd, Poole:2012ex}.

In this work we investigate \hi\ IM's suitability to probe scales around the turnover feature in the power spectrum. We examine the pathfinder survey MeerKAT's capability to detect a turnover, and also a more advanced SKAO survey's capability to constrain the precise position of the turnover and how this can potentially probe cosmological information. The challenge for such an objective is the broadness of the turnover feature which, compared to the more defined BAO, makes tight constraints difficult. Furthermore, BAO are considered particularly robust to systematics \citep{Seo:2003pu}, and it is not clear if the turnover would be similarly resilient. The challenge of constraining the turnover is also exacerbated by the fact that the feature is on very large scales, making any statistically significant detection limited by cosmic variance and survey size. However, as discussed, \hi\ IM is expected to be well-suited to probing large scales. The uncertainty on \hi\ bias and abundance should be no issue, since on large linear scales, this will only affect the amplitude of the \hi\ power spectrum and not the scale of the turnover location. Many \hi\ IM observational effects should also be at their most minimal in the area of $k$-space around the turnover, as we will demonstrate. Lastly, since the turnover scale is linear at all redshifts, the potential for performing a model-independent fit to the turnover, uncomplicated by non-linear phenomena, is promising and something we investigate.

The paper is structured as follows; in \secref{sec:HIIMPower} we introduce our \hi\ IM power spectrum formalism along with the assumed specifications for the surveys we study. In \secref{sec:Turnover} we introduce the model independent fitting to the turnover along with some detection forecasts. We also propose a new methodology for constraining the turnover location. \secref{sec:SimulatedDataTest} presents the results from the tests on our simulated data. In \secref{sec:CosmofromTurnover} we speculate on some possibilities for using constraints on the turnover location for a standard-ruler distance measurement and cosmological parameter inference. We then finally conclude in \secref{sec:Conclusion}.

\section{\hi\ power spectrum \& IM surveys}\label{sec:HIIMPower}

\begin{figure*}
	\centering
  	\includegraphics[width=2.05\columnwidth]{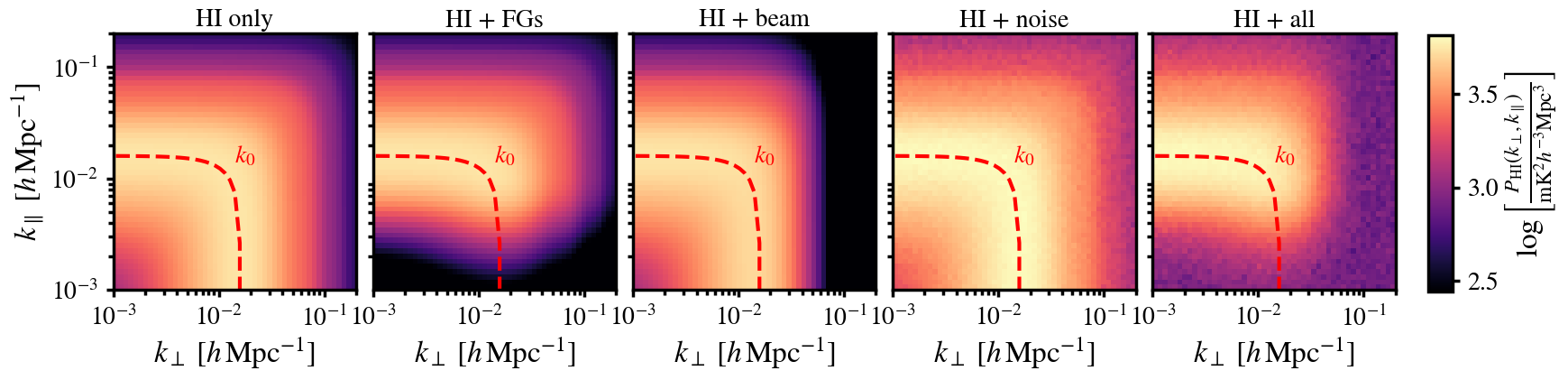}
    \caption{Robustness of the turnover scale $k_0$ to intensity mapping observational effects. We show a model \hi\ power spectrum decomposed into 2-dimensional modes perpendicular and parallel to the line-of-sight ($k_\perp,k_\parallel$). The highest intensity region is where the power spectrum peaks and is thus the turnover region. We mark this exact position with the turnover scale $k_0$ shown by the red dashed contour. In each panel we show the impact to the power spectrum from different observational effects including; signal loss due to foreground cleaning, a smoothing to the field caused by the telescope beam, and the presence of scale-independent thermal noise. \textcolor{black}{The final panel shows the combination from all observational effects.}}
\label{fig:TurnoverDetectionDemo}
\end{figure*}

\hi\ in the late Universe ($z\lesssim5$) is contained within galaxies, self-shielded from ionising radiation. By recording the unresolved emission of the redshifted 21cm spectral feature, we can construct 3D maps of \hi\ which have been shown to trace the underlying matter density \citep{Masui:2012zc,Anderson:2017ert,Wolz:2021ofa}. Measuring 2-point clustering statistics in 3D Fourier space for \hi\ IM will estimate a \hi\ power spectrum which will be described by
\begin{equation}\label{eq:HIpowerspec}
    P_\hi(k,\mu,z) = \overline{T}^2_\hi(z) \left[b_\hi(z) + f(z)\mu^{2}\right]^{2} P_\text{m}(k,z) \,,
\end{equation}
where $\overline{T}_\hi$ is the mean \hi\ temperature of the field, $b_\hi$ is the linear bias, and $P_\text{m}$ is the matter power spectrum. The $f\mu^2$ term is implemented to model the anisotropies from linear redshift-space distortions (RSD) \citep{Kaiser:1987qv}, where $f$ is the linear growth rate of structure. This scale-independent term will act as a boost to the amplitude of the power. On the highly linear scales we focus on in this paper, it is reasonable to assume all biasing terms in \autoref{eq:HIpowerspec} are scale independent, thus a measurement of turnover scale in the \hi\ power spectrum, should directly map to the turnover scale in the matter power spectrum $P_\text{m}$.

For the \hi\ bias we extrapolate a model based on hydrodynamical simulations \citep{Villaescusa-Navarro:2018vsg}
\begin{equation}\label{eq:HIbias}
    b_\hi(z) = 0.842 + 0.693z - 0.046z^2\,.
\end{equation}
The mean \hi\ temperature is related to \hi\ density abundance by \citep{Battye:2012tg}
\begin{equation}\label{eq:TbarModelEq}
    \overline{T}_\hi(z) = 180\Omega_{\hi}(z)h\frac{(1+z)^2}{H(z)/H_0} \, {\text{mK}} \, ,
\end{equation}
where for $\Omega_\hi$ we utilise an analytical function adopted in \citet{Pourtsidou:2016dzn,Bacon:2018dui}
\begin{equation}
    \Omega_\hi(z) = 0.00048 + 0.00039z - 0.000065z^2
\end{equation}
which is consistent with the latest \hi\ constraints \citep{Masui:2012zc,Crighton:2015pza,Wolz:2021ofa}.

The model in \autoref{eq:HIpowerspec} currently assumes there are no additional observational effects acting to distort the \hi\ power spectrum measurement, which is certainly not the case in reality. When using IM there are a number of effects to consider. We outline the most dominant below:

\textbf{21cm foregrounds}: Since IM records all unresolved diffuse emission in a particular frequency, it will not only capture the cosmological signal from extra-galactic \hi. There are several other astrophysical processes that emit radiation in the same frequency ranges, most of which dominate over the inherently weak cosmological \hi\ signal. The most dominant sources are synchrotron and free-free emission from within our own Galaxy, and also extra-galactic point sources from objects such as active galactic nuclei \citep{Santos:2004ju,Alonso:2014sna}. These foregrounds therefore require removing in order to access the \hi\ signal required to probe cosmology. Assuming a well calibrated instrument, these foregrounds will exhibit a continuum-like smooth spectrum through frequency, in contrast to the \hi\, which, due to its clean 21cm spectral feature traces cosmological structure, is discrete with redshift (and therefore frequency).

An effective way to perform a blind foreground clean is to therefore apply a Principal Component Analysis (PCA) to the foreground contaminated data. Since the foregrounds are the dominant sources and very correlated through frequency, the majority of their contribution should be contained in a small number of principal components, which can then be removed \citep{Liu:2011ih,Masui:2012zc}. This blind foreground removal method has been shown to be effective in data \citep{Masui:2012zc,Anderson:2017ert} and simulations \citep{Alonso:2014dhk,Cunnington:2020njn}, but inevitably causes some signal loss in the \hi\ density field. This is typically concentrated in the small-$k_\parallel$ modes, most degenerate with the foregrounds. As demonstrated in previous work on simulations \citep{Cunnington:2020wdu,Soares:2020zaq}, a model for describing the signal loss impact can be given by the below function which the \hi\ power spectrum is multiplied by.
\begin{equation}\label{eq:FGdampmodel}
    B_\text{fg}(\boldsymbol{k}) = 1-\exp \left[-\left(k_\parallel / k_\parallel^{\mathrm{fg}}\right)^2\right]\,.
\end{equation}
This damps signal coming from large radial modes (i.e. small-$k_\parallel$ modes) and the strength of such damping is modulated by the parameter $k_\parallel^{\mathrm{fg}}$, where the larger this is, the larger the signal loss.

In \autoref{fig:TurnoverDetectionDemo} we demonstrate the potential impact from some \hi\ IM observational effects on the power spectrum, focusing on the turnover scale $k_0$, the position of which is shown by the red-dashed line. We have decomposed the power spectrum into contributions from modes perpendicular and parallel to the line-of-sight ($\boldsymbol{k}_\perp,k_\parallel$) to highlight the anisoptropy of some of these effects. In the far-left panel we show just the pure \hi\ power spectrum, then in subsequent panels, we add in models of observational effects to demonstrate their impact. We will introduce the noise and beam effects later in this section. The impact from the foregrounds is shown in the second panel and has been modelled by \autoref{eq:FGdampmodel} using a parameter of $k_\parallel^{\mathrm{fg}}=5{\times} 10^{-3}\,h/\text{Mpc}$. \textcolor{black}{Our choice for this parameter is fairly arbitrary but based on previous simulation-based results \citep{Cunnington:2020wdu}. It will also depend on the survey size and telescope calibration. A survey requiring a more aggressive foreground clean due to effects such as polarisation leakage, will suffer more signal loss and have a greater $k_\parallel^{\mathrm{fg}}$.} As we expect, this drastically damps power at small-$k_\parallel$, and will have a sufficient impact on the turnover scale. Correcting for this will therefore be important and appears to be the most dominant of observational effects, relevant for the turnover and we therefore discuss this much further later in the paper and incorporate it into our simulated data to provide a robust investigation.

\textbf{Single-dish telescope beam}: For single-dish IM surveys with the SKAO and its pathfinders, the intensity pattern observed will be quite broad in the main beam, which means the observed density field is effectively smoothed in directions perpendicular to the line-of-sight. The reason the beam is large for these surveys is due to the relatively small dish size $D_\mathrm{dish}{\sim}15\,\text{m}$, which directly impacts the full-width-half-maximum (FWHM) size of the central lobe as $\theta_{\mathrm{FWHM}} \approx c / (\nu\,D_{\mathrm{dish}})$ for observations at frequency $\nu$. By approximating the beam pattern as a single central lobe with a Gaussian distribution, the effect on the density field is a simple Gaussian smoothing to perpendicular scales given by
\begin{equation}
    B_{\mathrm{beam}}(\boldsymbol{k})=\exp \left[-\frac{1}{2} k_{\perp}^{2} R_{\mathrm{beam}}^{2}\right]\,,
\end{equation}
where $R_\text{beam}$ is the standard deviation of the central beam in physical units, so given as $R_\text{beam} = r(z)\theta_{\mathrm{FWHM}} /(2 \sqrt{2 \ln 2})$, where $r(z)$ is the comoving distance\footnote{We assume the Universe has zero curvature throughout this work.} to density fluctuations under observation. 

The impact from the beam on the large scale turnover feature is demonstrated in the third panel of \autoref{fig:TurnoverDetectionDemo} where we have used $R_\text{beam}=20\,\text{Mpc}/h$, approximately the size of the SKAO single-dish beam at $800\,\text{MHz}$. Whilst this suggests that the beam should not be of large concern when probing large scale features such as the turnover, we highlight that the beam size will vary with frequency and reach a much larger size for high-redshift SKAO observations. Whilst this still will not come close to drastically eliminating the turnover scales, it can have a sufficient damping to the power spectrum such that the apparent turnover scale $k_0$ is shifted, thus returning biased measurements. Hence, careful calibration and modelling will be required in these cases for the purpose of precision cosmology. 

\textbf{Instrumental noise and RFI}: Since the \hi\ IM strategy involves integrating all signal down to the faintest emitters, shot-noise should be a sub-dominant component in the data \citep{Spinelli:2019smg}. However, thermal fluctuations in the electronics of the telescope cause instrumental noise which manifests as Gaussian random fluctuations in the maps. This can potentially be a significant component in the data, especially for early pathfinder surveys. However, a well controlled system temperature can limit this impact as can increasing the survey's observation time which will cause the Gaussian mean-centred noise to compound and thus reduce in overall amplitude.

Following \citet{Santos:2015gra} and the radiometer equation we can define the expected noise temperature rms as a function of frequency $\nu$ to be given by
\begin{equation}\label{eq:noise}
    \sigma_\text{N}(\nu) = \frac{T_\text{sys}(\nu)}{\sqrt{2\,\delta\nu\, t_\text{p}}}\,,
\end{equation}
where $T_\text{sys}$ is the total system temperature, $\delta\nu$ is the frequency channel width and $t_\text{p}$ is the time per pointing of the telescope. For the time per pointing, we will assume each pixel is $1/3$ of the beam size $\theta_\text{FWHM}$, which is approximately consistent with the MeerKAT pilot survey data in \citet{Wang:2020lkn}. From this the number of pointings required to fill a certain target survey size $A_\text{sky}$ can be calculated, and then the total observation time $t_\text{obs}$ shared among each pointing. Thus we define the time per pointing as
\begin{equation}\label{eq:timeperpointing}
    t_\text{p} = N_\text{dish}\, t_\text{obs}\left(\theta_{\mathrm{FWHM}}/3\right)^{2} / A_\text{sky}\,,
\end{equation}
where we have also multiplied through by the number of dishes $N_\text{dish}$ to account for the fact that the surveys are operating in single-dish mode, with each dish in the array making its own auto-correlation contribution.

For the system temperature in \autoref{eq:noise}, we follow the definition in \citet{Bacon:2018dui}
\begin{equation}
    T_\text{sys}(\nu) = T_{\mathrm{rx}}(\nu)+T_{\mathrm{spl}}+T_{\mathrm{CMB}}+T_{\mathrm{gal}}(\nu)\,,
\end{equation}
where the contribution from spill-over is $T_{\mathrm{spl}}=3\,\text{K}$, the background contribution from the cosmic microwave background (CMB) is $T_{\mathrm{CMB}}=2.73\,\text{K}$ and the contribution from our own Galaxy is $T_{\mathrm{gal}} =  25\,\mathrm{K}(408\,\mathrm{MHz} / \nu)^{2.75}$. For the receiver temperature, we use a slightly modified version which has been tuned to better match recent observations in the MeerKAT pilot survey \citep{Wang:2020lkn}, given by
\begin{equation}
    T_{\mathrm{rx}}(\nu)=7.5\,\mathrm{K}+10\, \mathrm{K}\left(\frac{\nu}{\mathrm{GHz}}-0.75\right)^{2}\,.
\end{equation}
If assuming a uniform Gaussian instrumental noise is present in the IM, the contribution to the power spectrum will be an additive component $P_\text{N}(\nu) = V_\text{cell}\,\sigma_\text{N}(\nu)$, where $V_\text{cell}$ is the volume of the survey's voxels. We demonstrate this in the fourth panel of \autoref{fig:TurnoverDetectionDemo}, where the additive contribution from the noise is clearly visible on small-scales. In this toy example we have used an unrealistically high level of noise to demonstrate the effect. Even in this case though, the turnover scales are the most robust to the noise since the peak power amplitude around these scales is the most likely to be large enough to dominate over additive contributions such as this instrumental noise. 

A further important noise-like component to consider comes from Radio Frequency Interference (RFI). These terrestrial signals can be unpredictable in occurrence and when present mostly dominate all other components. This is typically avoided with a rigorous flagging scheme whereby data is methodically checked at different stages, initially in the raw time-ordered data, deleting particular time chunks and channels where clear signs of RFI are present. This can be quite an aggressive process and in early pilot surveys, the majority of data ends up being flagged \citep{Wang:2020lkn}. The hope is this process will become more efficient and scanning strategies can become more sophisticated to avoid such contamination. Since this is currently poorly understood and difficult to simulate or model, we do not include it in this investigation and assume RFI has either been completely flagged, or residuals are minimal enough, not to impact the turnover scales.\newline

\noindent In this study to maintain a consistent approach across all forecast surveys (discussed in the following sub-section), we adopt the below formalism to define our power spectrum binning strategy and error estimation. We define the minimum scale probed by a survey as
\begin{equation}\label{eq:kmin}
    k_{\min} = 2\pi \Big/ \sqrt{l_\text{x}^2 + l_\text{y}^2 + l_\text{z}^2}\,,
\end{equation}
where $l_\text{x}$, $l_\text{y}$ and $l_\text{z}$ are the physical dimensions of the survey, which throughout this work we assume to be a simple cuboid. This simplification allows us to avoid the complication of wide-angle effects \citep{Castorina:2017inr,Blake:2018tou} which will be a challenge for large sky surveys such as those done with \hi\ IM. For our Cartesian grid we set $l_\text{z}$ as the comoving distance between the minima and maxima redshift range of the survey. We then define $l_\text{x}$ and $l_\text{y}$ such that they approximate the target angular sky coverage at the median redshift $z_\text{eff}$ i.e. $l_\text{x}=l_\text{y}=r(z_\text{eff})\sqrt{A_\text{sky}}$.

We use $k_{\min}$ to define the bin widths as $\Delta k= 2k_{\min}$. $k_{\max}$ can be chosen for optimal results and for this work a low value is preferred since little gain is made by including high-$k$ modes for evaluating the turnover, especially as these will start to carry more non-linear effects and influence from the large IM beam. The error associated with a power spectrum measurement can be estimated with
\begin{equation}\label{eq:Pkerr}
    \delta P_\hi(k) = \frac{P_\hi(k) + P_\text{N}}{\sqrt{N_\text{modes}(k)}}\,,
\end{equation}
where $N_\text{modes}$ is the number of modes in each $k$-bin calculated. We refer the reader to \citet{Blake:2019ddd} for a dedicated discussion on optimal power spectrum and error estimation. $N_\text{modes}$ is based on the survey size and analytically given by
\begin{equation}
    N_\text{modes}(k) = V_{\text{sur}} \frac{4\pi k^2 \Delta k}{(2\pi)^3}\,,
\end{equation}
where $V_{\text{sur}}$ is the total volume of the survey. $P_\text{N}$ in \autoref{eq:Pkerr} is the noise power spectrum contribution and determined using \autoref{eq:noise}. Since we are using very deep frequency ranges in this study, we calculate $\sigma_\text{N}$ at each frequency in the particular survey range, and use this to randomly generate a Gaussian noise map. The Gaussian noise at all frequencies are stacked into the full frequency range data cube which we then measure the power spectrum for to ensure a robust model of $P_\text{N}$, which is then used for the error estimation.

The values for the parameters relevant to the equations introduced in this section are dependent on the particular \hi\ IM survey being studied. We introduce the surveys we consider in this work in the following section.

\subsection{Large sky surveys with \hi\ intensity mapping}

To provide a wide range of forecasts for turnover constraints, we opt to study both an advanced final SKAO 197 dish survey and also a nearer-term pathfinder survey from SKAO's precursor MeerKAT, which is a smaller 64 dish array eventually to be combined into the final SKAO. The MeerKAT telescope is already operational and has successfully demonstrated calibration of the single-dish mode intensity mapping technique for an array of dishes \citep{Wang:2020lkn}. Furthermore, it is already beginning to deliver science results \citep{Irfan:2021xuk}. There is a plan for a MeerKAT Large Area Synoptic Survey (MeerKLASS) \citep{Santos:2017qgq} and it is this which we chose as our nearer-term survey and investigate if this will be capable of detecting a turnover in \hi\ power spectrum. MeerKLASS is planned to be a 4,000 hour survey covering 4,000$\,\text{deg}^2$ and can be performed in two different frequency bands; L-band ($0<z<0.58$) and UHF-band ($0.4<z<1.45$). We summarise the details for the MeerKLASS surveys in \autoref{tab:SurveyTable} for both frequency bands. To investigate more futuristic possibilities with \hi\ IM, we also include the planned wide Band 1 survey using SKA-MID, which will have 10,000 hours of observation, covering 20,000$\,\text{deg}^2$ across redshift $0.35<z<3$ \citep{Bacon:2018dui}. The details of which are also outlined in \autoref{tab:SurveyTable}.

\begin{table}
    \setlength{\tabcolsep}{4.5pt}
	\centering
	\begin{tabular}{lcccc} 
		\toprule
		 & & \multicolumn{2}{c}{\textbf{MeerKLASS}}  & \textbf{SKA-MID}\\
		\textbf{Survey Parameters} & & L-band & UHF-band & Band 1 \\
		\toprule
        Bandwidth [MHz] & $\nu_\text{min}$ &  900 & 580 & 350\\
         & $\nu_\text{max}$ & 1185$^*$ & 1000 & 1050 \\
    	\midrule
        Redshift range & $z_\text{min}$ & 0.2 & 0.4 & 0.35 \\
         & $z_\text{max}$ & 0.58 & 1.45 & 3 \\
		\midrule
        Effective redshift & $z_\text{eff}$ & $0.39$ & $0.925$ & $1.675$ \\
        Sky area [deg$^2$] & $A_\mathrm{sky}$ & 4,000 & 4,000 & 20,000\\
        Sky fraction & $f_\mathrm{sky}$ & $0.10$ & $0.10$ & $0.48$ \\
        Channel width [MHz] & $\delta\nu$ & $0.2$ & $0.2$ & $0.2$ \\
        Observation time [hrs] & $t_\mathrm{obs}$ & 4,000 & 4,000 & 10,000 \\
		Number of dishes & $N_\mathrm{dish}$ & $64$ & $64$ & $197$ \\
        Dish diameter [m] & $D_\mathrm{dish}$ & $13.5$ & $13.5$ & $15$\\
        Beam size (at $z_\text{eff}$) [deg] & $\theta_\mathrm{FWHM}$ & $1.5$ & $1.7$ & $2.6$ \\
        \textcolor{black}{Max scale $[h/{\rm Mpc}]{\times 10^3}$} & \textcolor{black}{$k_{\min}$} & \textcolor{black}{$3.3$} & \textcolor{black}{$1.6$} &
        \textcolor{black}{$0.53$} \\
        \bottomrule
	\end{tabular}
    \caption{Specifications for the MeerKLASS L-band, UHF-band, and SKA-MID Band 1 surveys \citep{Santos:2017qgq,Bacon:2018dui}. 64 of the SKA dishes will be the existing MeerKAT dishes that have $D_\mathrm{dish}=13.5\,\text{m}$, but for simplicity we make the approximation that all dishes have the same $15\,\text{m}$ diameter. $^*$The L-band actually extends up to ${\sim}1700\,\text{MHz}$ but we impose a low redshift cut at $z=0.2$ ($\nu_{\max}{\sim}1185\,\text{MHz}$) for our data.}
    \label{tab:SurveyTable}
\end{table}

Throughout the paper we assume that we are able to directly probe the entire redshift range from these surveys without re-binning into smaller bins. This allows for an optimal signal-to-noise on turnover constraints but is a demanding requirement since there will be significant cosmological evolution in the redshift ranges we consider. However, a similar challenge is posed in other experiments aiming to maximise signal-to-noise of large-scale probes \citep[e.g. see the discussion in][in the context of BAO]{Zhu:2014ica}. Current galaxy analyses probing primordial non-Gaussianity on ultra-large scales adopt similar approaches of very deep redshift bins \citep[e.g.][]{Castorina:2019wmr,Rezaie:2021voi}. Redshift weighting schemes can be constructed for specific probes \citep[e.g.][]{Ruggeri:2016mac,Mueller:2017pop} which allow for large bins to be used. A similar technique, in principle, could be developed for \hi\ IM measurements of ultra-large scales and even optimised for the turnover scales. We delay this technical task for future work, and here assign an effective median redshift $z_\text{eff}$ to all our surveys on which the underlying fiducial cosmology is based.

\begin{figure}
	\centering
    \includegraphics[width=\columnwidth]{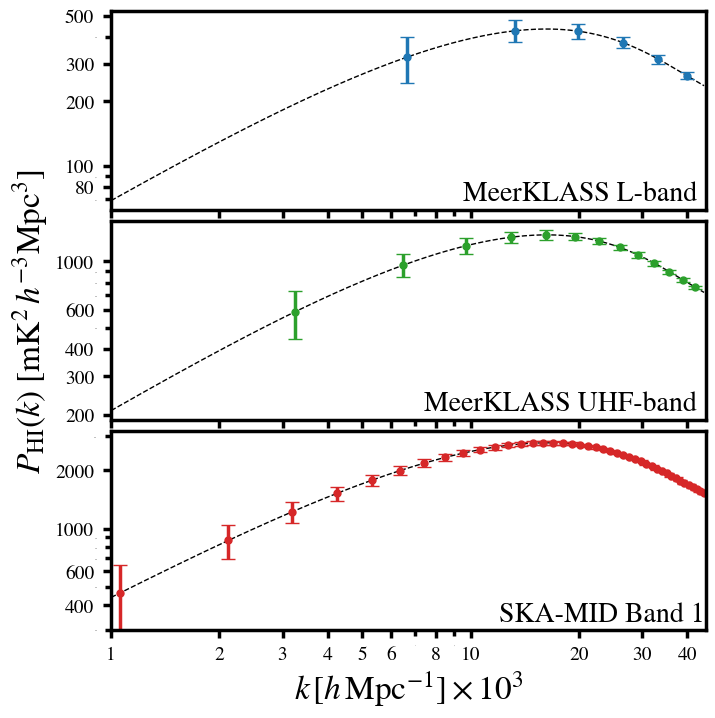}
    \caption{Synthetic power spectra data around the turnover region with upcoming \hi\ intensity mapping surveys. Data points, forecast error-bars, and $k_{\min}$ limitations have been determined following the formalism laid out in \secref{sec:HIIMPower}.}
    \label{fig:SurveyPks}
\end{figure}

To provide an idea for the power spectra constraints obtainable from the \hi\ IM surveys, we have plotted synthetic data points in \autoref{fig:SurveyPks}. These are simply generated from the model in \autoref{eq:HIpowerspec} and following the $k$-binning and error formalism outlined in the previous section. This simple demonstration already provides an idea for each surveys capability to detect and constrain the turnover position. It seems the MeerKLASS L-band survey may struggle to return a statistically significant detection. Results should improve in UHF band despite this being the same survey just at a different frequency range. The reason for this is because the UHF band probes a higher redshift range, which despite being the same angular size at the L-band, covers a wider and deeper physical volume, allowing for larger modes to be measured. As expected the much larger survey with the SKAO (bottom panel of \autoref{fig:SurveyPks}) provides a high chance of turnover detection, demonstrating excellent potential for signal-to-noise. In the following section we will introduce the formalism with which to evaluate the possibilities of detecting and constraining the turnover in a more quantitative manner.

\section{Turnover detection \& constraints with IM}\label{sec:Turnover}

The aim of this work is to evaluate if the data returned from the \hi\ IM surveys outlined in \autoref{tab:SurveyTable} will reveal a turnover feature in their measured power spectra with statistical significance. A quick glance at the forecast data points from \autoref{fig:SurveyPks} suggests that a turnover should be easily identified in the SKAO survey, but it is less clear for MeerKLASS. We therefore require a methodical way to quantitatively evaluate if a turnover is statistically present in the data. 

In this section we begin to construct a process for evaluating a turnover, and present results for various cases on the strength of a detection. We do this for synthetic data such as that presented in \autoref{fig:SimulationPks}. In \secref{sec:SimulatedDataTest} we extend this to more realistic simulated data which includes foreground contamination, which for now we do not consider.

\subsection{Model-independent power spectrum fitting}

To quantitatively analyse the turnover in the \hi\ power spectrum, we adopt a model-independent approach to estimating the turnover scale $k_0$ as done in \citet{Blake:2004tr}, \citet{Poole:2012ex} and more recently in \citet{Pryer:2021cut}. We fit the measured scales around the turnover to the parabolic model given by
\begin{equation}\label{eq:ParabolaFit}
    \log _{10}\left(\frac{P_\hi(k)}{[{\rm mK}^2 h^{-3}{\rm Mpc}^3]}\right)=\left\{\begin{array}{ll}
    \log _{10}(P_{0})\left(1-\alpha x^{2}\right) & \quad k<k_{0} \\
    \log _{10}(P_{0})\left(1-\beta x^{2}\right) & \quad k \geq k_{0}\,,
    \end{array}\right.
\end{equation}
where
\begin{equation}
    x = \frac{\log_{10}(k/[h/\text{Mpc}]) - \log_{10}(k_0/[h/\text{Mpc}])}{\log_{10}(k_0/[h/\text{Mpc}])}\,.
\end{equation}
Here there are four free parameters $\Theta=\left\{P_0,\alpha,\beta,k_0\right\}$. $P_0$ is the unit-less peak amplitude of the power spectrum. $k_0$ is the position of the peak i.e. the turnover location. $\alpha$ and $\beta$ control the parabolic decrease of the power spectrum either side of the turnover. This model can therefore be used to determine if a turnover is present in the data by analysing the constraints obtained on the value of $\alpha$. If the confidence intervals on $\alpha$ suggests it is greater than zero, then this is evidence for the power spectrum peaking and turning over. However, $\alpha{<}0$ suggests the model is not preferentially fitting a peaked parabola suggesting no turnover is present, thus rendering any constraints on the turnover scale $k_0$ unreliable.

As discussed in \citet{Poole:2012ex}, at scales smaller than the turnover ($k>k_0$), the power spectrum has a changing logarithmic slope. Therefore this simple  logarithmic parabola will only provide a reasonable fit to a real power spectrum at a constricted scale range. Including data points up to a high $k_{\max}$ will also begin to include BAO and non-linear contributions not captured by the model, and potentially cause large biases. In this work, we impose a $k_{\max}=0.05\,h/\text{Mpc}$ for all our surveys to avoid this issue. This has the additional benefit for \hi\ IM of avoiding regions of $k$-space more likely to be affected by telescope beam issues, and noise-like contributions, as discussed in the previous section and presented in \autoref{fig:TurnoverDetectionDemo}.

We obtain constraints on the model parameters by exploring their posterior distribution through a likelihood analysis of the data. The  likelihood $\mathcal{L}$ is given by
\begin{equation}\label{eq:loglikelihood}
    -2\ln\mathcal{L}(\Theta) = \Delta \mathbf{P}(\Theta)^T\hat{\mathbf{C}}^{-1}\Delta\mathbf{P}(\Theta)\,,
\end{equation}
where $\Delta\mathbf{P}$ is the difference between the measured data points from the \hi\ power spectrum for the chosen $k$-range and those returned by the model with given parameters from the vector $\Theta$. Throughout we will assume a perfectly uncorrelated covariance matrix $\mathbf{C}$ whose diagonal elements are given by the error estimations given in \autoref{eq:Pkerr}.

We will initially fit the parabola model to the synthetic data presented in \autoref{fig:SurveyPks}, and use a Monte Carlo Markov Chain (MCMC) to explore the parameter space. This will allow us to test the model and choice of priors, as well providing a forecast for each survey's capability for detecting and constraining the turnover. In \secref{sec:SimulatedDataTest} we will move the investigation to more robust simulated data. For all the MCMC analysis in this work we use the publically available code \texttt{zeus}\footnote{\href{https://github.com/minaskar/zeus}{github.com/minaskar/zeus}} \citep{Karamanis:2020zss,Karamanis:2021tsx}.

In the analysis we use a fairly wide $k_0$ prior of $0.005<k_0<0.025\,h/\text{Mpc}$. Anything outside this range would be clearly nonphysical. The $k_0$ value for our fiducial cosmology is $k_0=0.0163\,h/\text{Mpc}$ at the effective redshift for the SKAO survey. Note that $k_0$ only varies slowly with redshift and has a sub-percent change between the low-redshift L-band data and the higher redshift SKAO Band 1 data. To make comparisons more convenient, we therefore often assume the same fiducial $k_0$ for all surveys. 

For $\alpha$ we place a $-3<\alpha<5$ prior. This allows for possible $\alpha{<}0$ values to be returned which would confirm that a turnover detection has not been made.  We place a flat positivity prior on $\beta$. We find no prior is needed for $P_0$ which is beneficial given how unconstrained the scaling parameters for \hi\ bias and abundance ($b_\hi$,$\Omega_\hi$) are. For this model we are only interested in the values of $\alpha$ (to confirm a turnover) and $k_0$ (to constrain the turnover location), thus we marginalise over the $P_0$ and $\beta$ parameters. But in all cases we check appropriate convergence in the MCMC chains has been achieved.

\subsection{Will MeerKAT detect a turnover?}\label{sec:WillMeerKATturnover}

\begin{figure*}
	\centering
    \includegraphics[width=2\columnwidth]{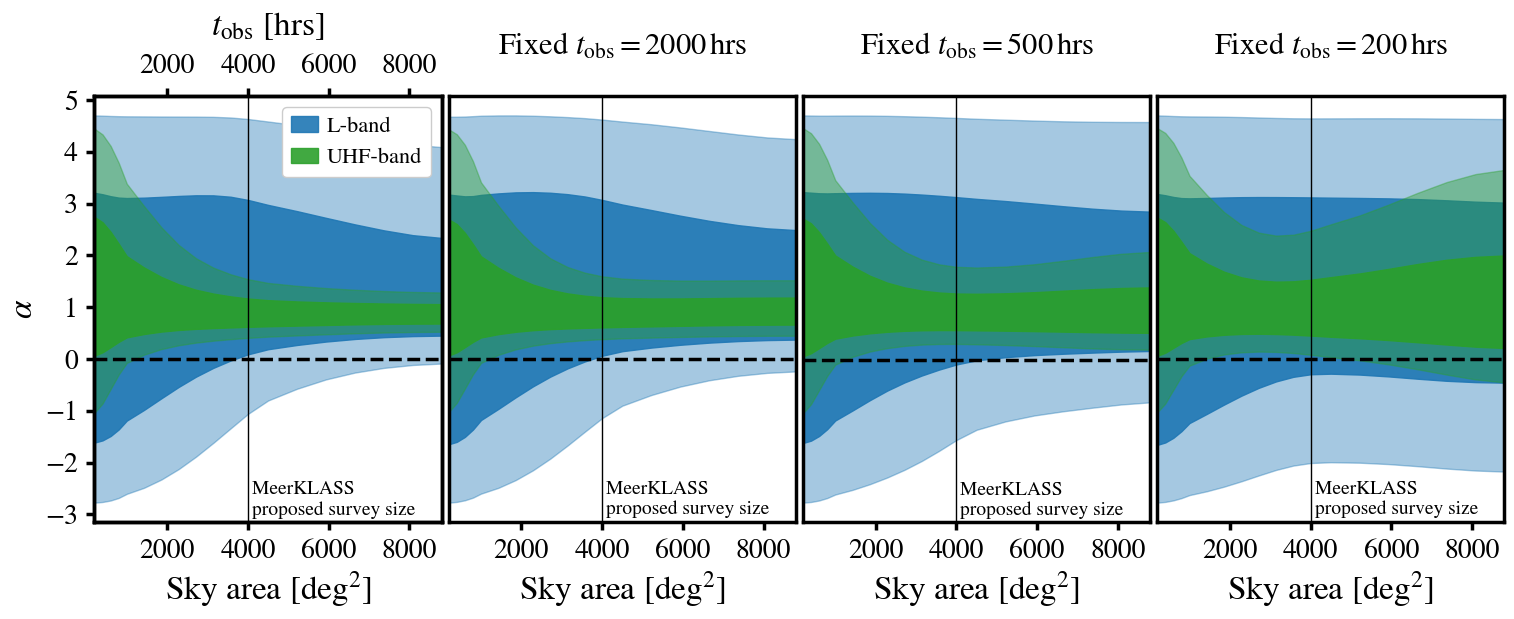}
    \caption{Sky area required to achieve detection ($\alpha{>}0$) of the turnover feature in the power spectrum for a MeerKAT survey in L-band and UHF-band. Shaded areas represent the $1\sigma$ and $2\sigma$ confidence regions on $\alpha$ measurements. Thin black vertical line indicates the sky area currently proposed for a MeerKLASS survey \citep{Santos:2017qgq}. \textit{Far-left panel} assumes a constant time per pointing ($t_\text{p}$) for each area, hence why the total observation time scales proportionally with sky area (shown by top $x$-axis). This keeps noise at a constant level as area changes. \textit{Other panels} show different cases of fixed total observation time (displayed in panel titles), all lower than that currently proposed for MeerKLASS (4,000$\,$hrs). In these panels, noise increases with increasing area, which is why constraints at large areas begin to worsen.}
    \label{fig:SkyArea_alpha}
\end{figure*}

Here we begin to quantitatively investigate if detecting the turnover feature will be possible with a near-term \hi\ IM survey such as MeerKLASS. We begin by examining the approximate area required for a successful detection. Whilst 4,000$\,\text{deg}^2$ is the proposed survey size for MeerKLASS, there is already smaller pilot survey data available \citep[e.g.][]{Wang:2020lkn} and it is likely further intermediate observations will be performed on increasingly larger areas before the full proposed MeerKLASS survey is conducted. Thus, gaining an idea for what particular survey volume is required for a turnover detection should be useful for survey planning. \autoref{fig:SkyArea_alpha} demonstrates the expected $1\sigma$ and $2\sigma$ confidence intervals on the parameter $\alpha$ for MeerKLASS surveys in both its frequency bands for different sky area coverage. This plot was produced by generating synthetic model data, varying the survey area each time and defining the $k$-binning and error bars based on \autoref{eq:kmin} and \autoref{eq:Pkerr}. The far-left panel uses a constant time per pointing ($t_\text{p}$) defined by the $t_\text{p}$ of the full MeerKLASS survey, which ensures noise (\autoref{eq:noise}) does not change as we vary survey area. Considering \autoref{eq:timeperpointing}, we can see that this means observation time ($t_\text{obs}$) scales proportionally to the sky area and both are numerically equivalent given the MeerKLASS proposal of 4,000$\,$hrs for a 4,000$\,\text{deg}^2$ survey. The other three panels use a fixed total observation time displayed in the panel title.

We can immediately see how a turnover ($\alpha{>}0$) detection in MeerKAT's L-band will be challenging and may require going beyond the proposed MeerKLASS survey size to guarantee an above $1\sigma$ detection. This is due to the lower redshift of the L-band and more limited redshift range which does not allow for a sufficiently large volume to be covered. This means cosmic variance begins to dominate the error budget on the largest scales around the turnover peak for the L-band as shown in \autoref{fig:SurveyPks}. With an insufficient resolution of modes around these scales, these results forecast only a mild $0.94\sigma$ statistical significance for $\alpha{>}0$ with MeerKAT's L-band.

In contrast, MeerKAT's UHF-band performs much better and should be capable of a $3.1\sigma$ turnover detection with the proposed 4,000$\,\text{deg}^2$ survey. Even if using lower areas, which may be more readily available, a detection should still be possible. The UHF-band which will be at higher and deeper redshifts relative to the L-band, therefore covers larger volumes rendering it a more promising probe of the turnover scales.

Detecting a turnover with \hi\ IM pathfinder surveys would be a significant achievement. \hi\ IM is yet to successfully achieve detection of a cosmological power spectrum in auto-correlation. A reason why this is so challenging is due to the high-levels of noise and residual systematics in the data. These generally cause additive biases and will boost the amplitude of power in an auto-correlation. Attempts to clean or model these systematics  are difficult since it is hard to know when the amplitude of the power has reduced enough such that the remaining contribution is cosmological signal, especially since there is large uncertainty on the true amplitude of the \hi\ power spectrum due to the unconstrained bias $b_\hi$ and abundance $\Omega_\hi$. The benefit of detecting a cosmological feature is that more confidence can be placed in the detection of cosmological signal if the target feature appears and is at the right location. In other words, identifying a turnover in power at the appropriate scale would be a cosmological smoking gun and makes a strong case for cosmological detection. Whereas merely fitting a featureless\footnote{The power spectrum will begin to display BAO wiggles at $k>k_0$, however the large telescope beam coupled with high noise could render such features undetectable, as discussed in the introduction.} sloping power spectrum at smaller scales is open to scrutiny concerning how much contribution comes from signal or systematic to reach an uncertain model amplitude.

\begin{figure}
	\centering
  	\includegraphics[width=\columnwidth]{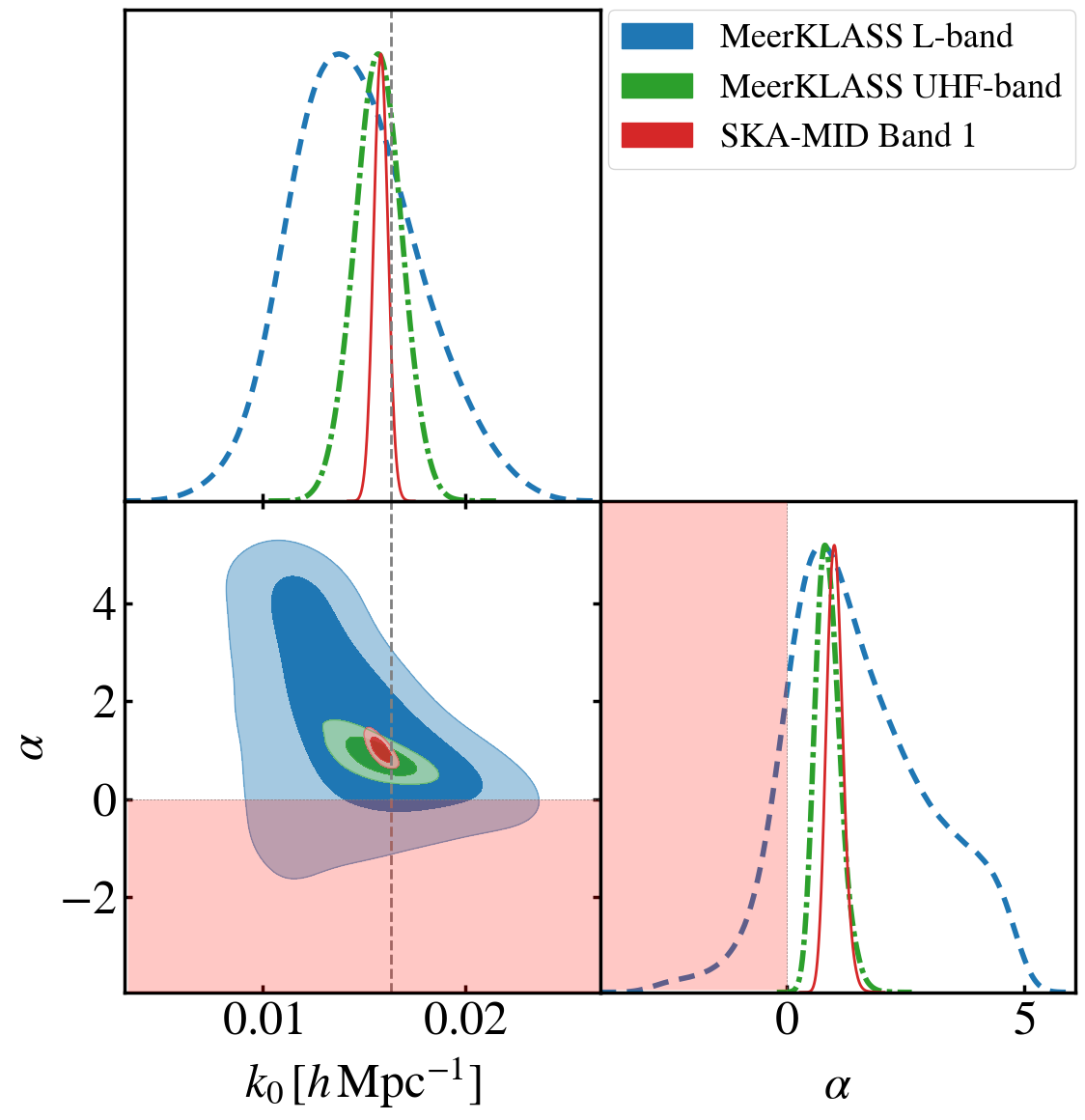}
    \caption{Posterior distributions for $k_0$ and $\alpha$ parameters for forecast \hi\ intensity mapping data using the parabola fitting technique. \textit{Pink-shaded} region marks the $\alpha{<}0$ area of parameter space which signifies an unsuccessful detection of a turnover feature. \textit{Grey-dashed} vertical line marks the fiducial turnover scale $k_0$. The MeerKLASS surveys appear consistent with the fiducial $k_0$ and constraints improve with the volumes of the survey. A scale cut was needed on the SKAO data at both small and large-$k$. Even with this scale cut, a slight bias is still evident in $k_0$, which we discuss in the text.}
\label{fig:MKalphk0posterior}
\end{figure}

To further demonstrate the potential constraints around the turnover scales we can extend the analysis of the synthetic model data and examine the posterior distribution in the $k_0 - \alpha$ plane. This is presented in \autoref{fig:MKalphk0posterior}. As expected, we see a consistency with \autoref{fig:SkyArea_alpha} and see a clear preference for positive $\alpha$ in the posterior for all surveys. Only the MeerKLASS L-band has any part of its $1\sigma$ confidence interval in the $\alpha{<}0$ region of parameter space, shown by the shaded pink. 
For the MeerKLASS surveys, we see a good agreement with the fiducial $k_0$ shown by the vertical grey dashed line. We also include SKAO results in \autoref{fig:MKalphk0posterior} which demonstrate how tighter constraints will be achievable with future surveys. Interestingly, we see a slight bias appearing on the $k_0$ posterior for this SKAO data. Given this is still being run on the synthetic model data, we would expect a much more accurate recovery of the fiducial $k_0$. We also found this bias worsens if the error bars are artificially reduced. This motivated an improved model to achieve unbiased $k_0$ constraints and it is this which we discuss in the following \secref{sec:LogFit}.

We explicitly outline the turnover detection forecasts in \autoref{tab:OpticalComparison} for the MeerKAT surveys along with the SKAO result which as shown, achieves a near-certain turnover detection, as expected. We also show some forecasts for various galaxy surveys which we discuss next.

\subsubsection{Comparison with optical galaxy surveys}\label{sec:OptComparison}

To provide some context for the forecast turnover detection significance we provide some approximate comparisons with both current and future galaxy surveys conducted in the optical and near-infrared wavelength ranges. \textcolor{black}{Galaxy surveys will also be conducted with radio instruments such as the SKAO. However, most of the detected galaxies will be at low redshifts ($z\lesssim 0.4$) \citep{Bacon:2018dui} and are unlikely to provide competitive volumes to probe turnover scales. We therefore do not consider comparisons with galaxy surveys with radio telescopes.}

We present the assumed survey specification for different galaxy surveys in \autoref{tab:OpticalComparison} along with their forecast detection significance for the turnover ($\alpha{>}0$). We draw inspiration from literature on complete (Stage-III) galaxy surveys \citep[e.g.][]{eBOSS:2020yzd,DES:2021bpo} and future Stage-IV surveys \citep[e.g.][]{DESI:2016fyo,LSSTDarkEnergyScience:2012kar}. We use the exact same techniques for the \hi\ IM model forecasts we have presented so far, following the methodology for $k$-binning outlined in \secref{sec:HIIMPower}. One difference is the calculation of uncertainties. For the optical galaxies, we assume the errors associated with the power spectrum estimation is given by
\begin{equation}\label{eq:Pkgerr}
    \delta P_\text{gal}(k) = \frac{P_\text{gal}(k) + 1/\bar{n}}{\sqrt{N_\text{modes}(k)}}\,,
\end{equation}
where the galaxy number density $\bar{n}$, accounts for the shot-noise in the surveys. We also quote the model \hi\ IM forecasts in \autoref{tab:OpticalComparison} to provide easy comparison. As with the IM forecasts, for simplicity, all galaxy surveys assume a square footprint matching the quoted sky area.

\begin{table*}
    \setlength{\tabcolsep}{4.5pt}
	\centering
	\begin{tabular}{lccccccc}
		\toprule
		\textbf{\hi\ IM survey} & $z_{\min}$ & $z_{\max}$ & Area [$\text{deg}^2$] & \textcolor{black}{Volume [$({\rm Gpc}/h)^3$]} & $t_\text{obs}$ [hrs] & \textcolor{black}{$P_0/P_{\rm N}$} & $\alpha{>}0$ \\
		\midrule
        MK L-band & 0.2 & 0.58 & 4,000 & \textcolor{black}{1.3} & 4,000 & \textcolor{black}{60.3} & 0.94$\sigma$\\
        MK UHF-band & 0.4 & 1.45 & 4,000 & \textcolor{black}{10.8} & 4,000 & \textcolor{black}{41.1} & 3.1$\sigma$\\
        SKAO Band 1 & 0.35 & 3 & 20,000 & \textcolor{black}{221.6} & 10,000 & \textcolor{black}{7.3} & 13.1$\sigma$\\		\toprule
		\textbf{Galaxy survey} & $z_{\min}$ & $z_{\max}$ & Area [$\text{deg}^2$] & \textcolor{black}{Volume [$({\rm Gpc}/h)^3$]} & $N_\text{gal}$ & \textcolor{black}{$P_0/P_{\rm SN}$} & $\alpha{>}0$ \\
		\midrule
        Stage III spectro-$z$ & 0.6 & 1.1 & 4,000 & \textcolor{black}{4.7} & $500{\times}10^3$ & \textcolor{black}{1.9} &  0.87$\sigma$\\
        Stage III photo-$z$ & 0.2 & 1.05 & 5,000 & \textcolor{black}{7.0} & $200{\times}10^6$ & \textcolor{black}{700.3} & 2.3$\sigma$\\
        Stage IV spectro-$z$ & 0.4 & 1.6 & 14,000 & \textcolor{black}{46.8} & $21{\times}10^6$ & \textcolor{black}{7.2} & 5.9$\sigma$\\
        Stage IV photo-$z$ & 0.3 & 3 & 20,000 & \textcolor{black}{225.6} & $10{\times}10^9$ & \textcolor{black}{445.9} & 16.9$\sigma$\\
        \bottomrule
	\end{tabular}
    \caption{Statistical significance of turnover detections ($\alpha{>}0$) for \hi\ IM and a comparison with galaxy surveys in optical and near-infrared wavelengths. The \hi\ IM results are for the two MeerKLASS surveys and the SKA-MID Band1 survey. Full details for these are outlined in \autoref{tab:SurveyTable}. No consideration has been given to the impact from redshift uncertainties in the photo-$z$ surveys. The assumed specifications for the quoted galaxy surveys are discussed in \secref{sec:OptComparison}.}
    \label{tab:OpticalComparison}
\end{table*}

As discussed in the introduction, the main limitation from spectroscopic galaxy surveys is from the number of galaxies observed over a sufficiently large volume\textcolor{black}{, as shown by the high shot-noise relative to the turnover power amplitude, $P_0/P_{\rm SN}$, displayed in the penultimate column of \autoref{tab:OpticalComparison}}. The results suggest that the SKAO survey should at least double the turnover constraints obtained from a completed Stage-IV spectroscopic galaxy survey. Current Stage-III spectroscopic galaxy surveys can also be far surpassed by MeerKLASS. The photometric surveys provide more optimistic results, although it still appears MeerKLASS can surpass a current Stage-III survey. A caveat to the photo-$z$ results is that there is no inclusion for redshift uncertainties. In a similar way to how the IM beam will act to damp small-scale modes, photo-$z$ uncertainties should only cause some statistical errors on small-scales. This is unlikely to have a large impact on the large-scale turnover feature assuming systematic catastrophic errors are well controlled \citep{Blake:2004tr}.

\subsection{Logarithmic polynomial model: an improved fit}\label{sec:LogFit}

As shown by the results from using the parabola fitting technique in \autoref{fig:MKalphk0posterior}, a slight bias appears in the $k_0$ distribution for the SKAO data. We experimented with different scale cuts but were still unable to improve upon the results presented. This would potentially be an issue for future surveys looking to tightly constrain the turnover location and use it for probing other cosmological information (something we investigate later in \secref{sec:CosmofromTurnover}). The recent study in \citet{Pryer:2021cut} interestingly found a similar conclusion, using simulated data from galaxy lightcone mock surveys, produced from $N$-body simulations. They also found their recovered $k_0$ value, using the same model-independent parabola fit, also favoured a slightly biased smaller value of $k_0$ relative to their fiducial input.

There could be a number of reasons for the struggling performance of the parabola fit with larger volume data. One explanation could be due to the higher concentration of small-$k$ modes such data begins to probe. We tried to implement the extended model outlined in \citet{Poole:2012ex} which they applied to futuristic larger volume survey data. They argue that since the primordial power spectrum should begin to emerge in a survey’s largest modes, this should render the simple asymmetric parabola model insufficient. They add an additional degree of freedom to describe how quickly $P(k)$ asymptotes to a power law at $k<k_0$. We experimented with this method but still found it did not return results with a sufficient accuracy to recover unbiased $k_0$ constraints.

\begin{figure}
	\centering
  	\includegraphics[width=\columnwidth]{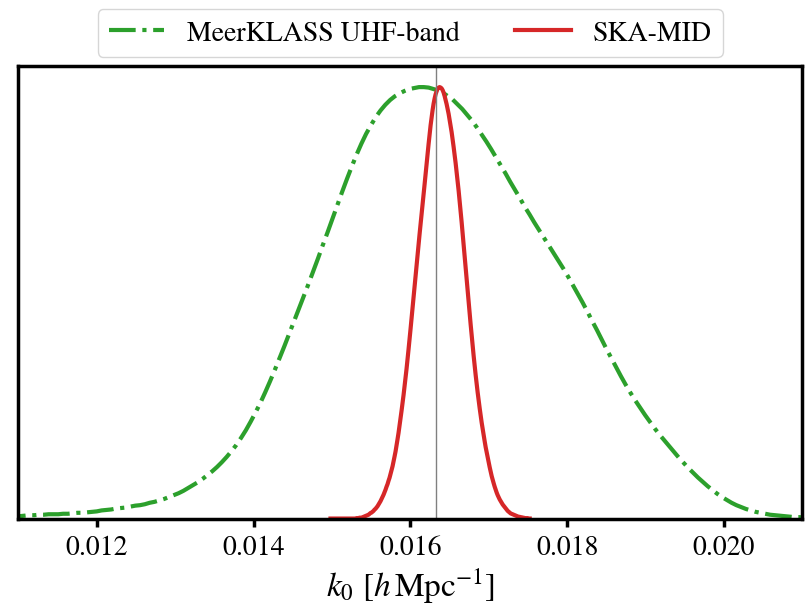}
    \caption{Demonstration of the unbiased constraints on the turnover scale $k_0$ possible with the logarithmic polynomial fit method. Unlike the parabola method used previously (e.g. in \autoref{fig:MKalphk0posterior}) this does not appear to bias the $k_0$ measurement relative to the fiducial value for the more constrained SKAO data.}
    \label{fig:k0histograms}
\end{figure}

We therefore propose an alternative approach involving fitting a simple polynomial to the logarithm of the power spectrum;
\begin{equation}\label{eq:PolyfitModel}
    \kappa^3\log _{10}\left(\frac{P_\hi(k)}{{\rm mK}^2 h^{-3}{\rm Mpc}^3}\right) = a_0 + a_1\kappa + a_2\kappa^2 + a_3\kappa^3 + a_4\kappa^4,
\end{equation}
where $\kappa=\log_{10}(k/[h/\text{Mpc}])$. As well as the previous peak-parabola fitting technique (\autoref{eq:ParabolaFit}), this model is also independent of cosmological inputs or model power spectra produced from Boltzmann solvers. It also has a very simple structure with a low number of free parameters. The logarithmic polynomial in \autoref{eq:PolyfitModel} does not however explicitly provide a scale for $k_0$. Yet by MCMC sampling with this model we can locate the turnover peak for each set of returned polynomial coefficients $\{a_0,a_1,a_2,a_3,a_4\}$ and build a posterior distribution for $k_0$. As with the previous parabola method, we estimate these posteriors using the likelihood defined in \autoref{eq:loglikelihood} but now with the parameter vector given by $\Theta=\{a_0,a_1,a_2,a_3,a_4\}$ which are input to the logarithmic polynomial model in \autoref{eq:PolyfitModel}. We found this method less susceptible to returning biased measurements of $k_0$ when extending it to the more constrained data from the SKAO survey. We also found the method to be far less sensitive to scale cuts, whereas the parabola technique could recover inaccurate results if $k_{\max}$ was too large, or $k_{\min}$ too small. 
We demonstrate the ability of the logarithmic polynomial fit method in \autoref{fig:k0histograms} where we are still using the synthetic \hi\ IM model data shown in \autoref{fig:SurveyPks}, but we are now able to recover more accurate measurement of $k_0$, in comparison to \autoref{fig:MKalphk0posterior} where the parabola fitting technique was used.

We emphasize a limitation of this logarithmic polynomial method is that it implicitly assumes a turnover is present in the data. If this is clearly the case, then it can be effectively used to return accurate constraints on the exact location of the turnover, $k_0$. This means the previous peak parabola fitting technique is still useful for evaluating if a turnover detection has been made. Indeed the parabola technique is likely to be more useful for pathfinder surveys in quantitatively establishing if a turnover in the data is present thus providing a cosmological smoking gun, as we discussed in \secref{sec:WillMeerKATturnover}. However, for more advanced surveys such as SKAO, with aspirations for conducting precision cosmology, the logarithmic polynomial method could provide a viable option for constraining $k_0$. \textcolor{black}{Furthermore, it could still be possible to quantitatively ascertain whether a turnover exists in the returned fitted polynomial. Simply identifying if the peak of the polynomial is not at the smallest-$k$ value would suggest a turnover is present. However, parameterising this and then evaluating a posterior for this turnover parameter is not something we have developed yet in this work.}

In the rest of this paper the focus will be on pursuing these tight constraints on $k_0$, examining how robust they are in simulated data with some observational effects, and also if there is potential for their constraints to be used for other cosmological parameter inference.

\section{Results from simulated data tests}\label{sec:SimulatedDataTest}

To extend our forecasts presented in \secref{sec:Turnover}, we look to conduct analysis on more robust simulated data. Real data from \hi\ IM will include some intrinsic scatter, especially on large scales due to irreducible cosmic variance, thus it is necessary to check our fitting models are robust to such variance. We also begin to include some testing of observational effects on a turnover measurement. 

Due to the limited success from the forecast turnover detection on the MeerKLASS L-band data, we do not include this in our $k_0$ estimation test since a constraint on this parameter is unreliable without a strong detection of a turnover. For the MeerKLASS UHF-band and SKAO surveys (details listed in \autoref{tab:SurveyTable}) we generate 20 lognormal \hi\ fields \citep{ColesLognormal1991} using the \hi\ power spectrum in \autoref{eq:HIpowerspec} as the input. We assume a Planck18 cosmology \citep{Aghanim:2018eyx} and use a CLASS Boltzmann solver \citep{Lesgourgues:2011re,Blas:2011rf} via \texttt{nbodykit}\footnote{\href{https://nbodykit.readthedocs.io/en/latest/index.html}{nbodykit.readthedocs.io}} \citep{Nbody_Hand} to generate the matter power spectrum. As discussed in \secref{sec:HIIMPower}, the dimensions of each simulation ($l_\text{x},l_\text{y},l_\text{z}$) is based on the survey in question, which we embed onto a grid of cells with size $n_\text{x}=n_\text{y}=n_\text{z}=256$.

Once a \hi\ field has been simulated using these steps, we run a power spectrum estimation on the field, using the same $k$-binning as discussed in \secref{sec:HIIMPower} and estimate the errors based on \autoref{eq:Pkerr}.

\subsection{Robust test of foreground contamination}\label{sec:FGtests}

The simulated data allows us to include realistic observational effects which threaten to bias a sensitive measurement of the turnover. As we demonstrated early in \autoref{fig:TurnoverDetectionDemo}, a likely observational effect to impact the turnover measurement will be signal loss caused by foreground cleaning. This is because foreground cleaning will distort the shape of the \hi\ power spectrum on large scales. If not corrected for this will drastically change the position of the turnover location, rendering any constraints on it biased and unreliable.

To investigate this we add onto each lognormal \hi\ simulation a realistic foreground map, which dominates the \hi\ amplitude. We then perform a PCA clean on the data which creates an accurate emulation of the signal loss which could be caused in real data. Left untreated, we found that the signal loss from foreground cleaning led to failures in the constraints on $k_0$ (shown by \autoref{fig:TransferFuncDemo}). Typically a foreground clean damps small-$k$ modes and this distorts the shape of the \hi\ power spectrum on large scales around the turnover. This shifts the peak position causing poor constraints on $k_0$.

To correct for the distorting effects caused by the foreground clean, we apply a foreground transfer function \citep{Switzer:2015ria,Cunnington:2020njn} as performed on real data analysis \citep[e.g.][]{Switzer:2013ewa,Wolz:2021ofa}. The foreground transfer function is constructed by injecting known mock data into the real data, then analysing the signal loss undergone in the mocks, using it to estimate the signal loss in the real data. The \hi\ power spectrum can therefore be corrected for. To demonstrate a robust and fair test of this process, we assume an incorrect cosmology for the injected mocks as could be the case in a real experiment. We thus use a cosmology based on the early WMAP5 results \citep{WMAP:2008lyn}. We discuss in more detail the foreground simulations and construction of the transfer function in \appref{app:FGsims}.

We find that using a transfer function to correct for signal loss appears to work effectively and recovers the shape of the \hi\ power spectrum, even where an incorrect cosmology is assumed for the mocks used in the transfer function construction. This can be seen in \autoref{fig:SimulationPks} where the data points show the averaged power spectrum for all 20 simulations with foreground contamination included. We see they are recovering the input \hi\ power spectrum model (black dashed line) even at the largest scales where the effects from foreground cleaning will be the most substantial. This is the case for both the MeerKAT and SKAO surveys. We also include some information on the $k_0$ constraints in \autoref{fig:SimulationPks} but disuss this in the following section.

The apparent success of a cosmological-independent foreground transfer function is an encouraging result in general for \hi\ IM experiments, but certainly requires a more involved dedicated study. It is highly likely that the nature of foreground contamination will be more complex when combined with other systematics, certainly in pathfinder survey data, which could render this a more complex task. We discuss this further in \appref{app:Systematics}. However, as an initial test, this appears to be a sufficient solution to what is likely to be one of the main distortions to the \hi\ power spectrum on the large scales around the turnover. 

\subsection{Constraints on the turnover scale $k_0$}

\begin{figure}
	\centering
  	\includegraphics[width=\columnwidth]{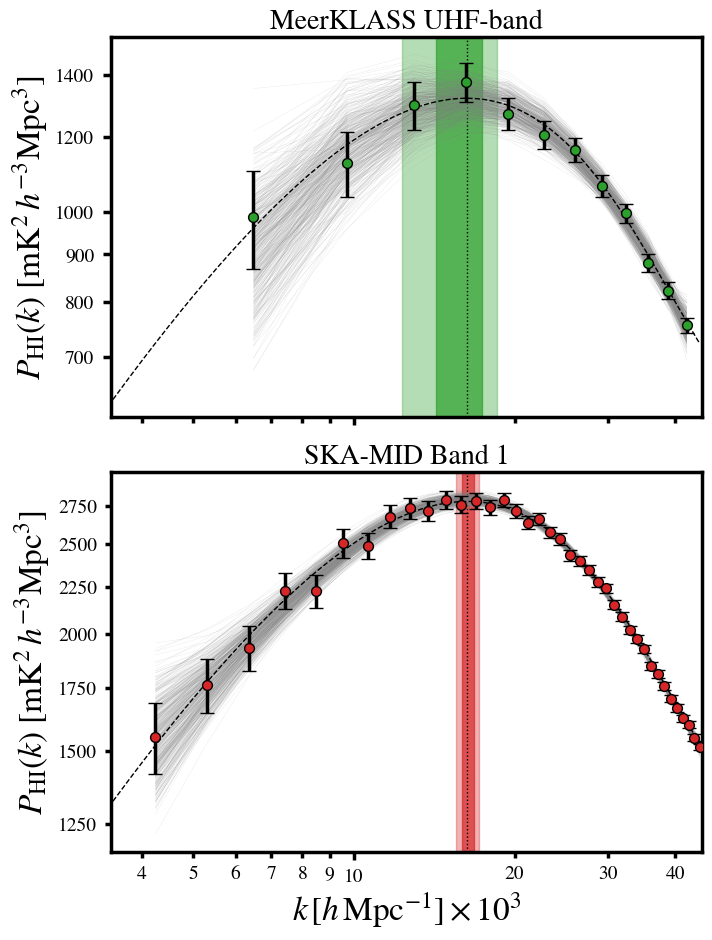}
    \caption{Results from simulated data tests inclusive of foreground contamination. The data points are the average from the 20 simulation realisations. Each simulation had a transfer function applied to correct for any signal loss from the foreground clean. The \textit{black-dashed} line represents the underlying input model for the simulations which the data measurements are recovering. The \textit{thin-grey} lines show some samples from the MCMC logarithmic polynomial fit which are used to measure the turnover scale $k_0$. The $1\sigma$ and $2\sigma$ confidence intervals on $k_0$ from this method are shown by the vertical shaded regions, with the \textit{black-dotted} vertical line showing the fiducial $k_0$.}
    \label{fig:SimulationPks}
\end{figure}

We fitted the logarithmic polynomial model to the averaged power spectra data points in \autoref{fig:SimulationPks} to test its constraining capability on the turnover scale $k_0$. To demonstrate the method working effectively, we plot some of the samples after burn-in from the MCMC analysis as thin-grey lines. These fit to the data points accurately and within the precision of the estimated error bars. It is from these that we ascertain a turnover position and construct a posterior distribution for $k_0$. We also present the confidence intervals from this posterior on \autoref{fig:SimulationPks} shown by the vertical shaded bars which mark the 1 and $2\sigma$ confidence regions. It is apparent that this method is doing a sufficient job at recovering an unbiased estimate for the fiducial $k_0$ (shown by the black vertical dotted line), which comfortably lies within the 1$\sigma$ interval for both MeerKAT's UHF band and the SKAO survey.

For the logarithmic polynomial fitting performed in \autoref{fig:SimulationPks}, as well as the $k_{\max}$ cut used in all previous analysis, we also applied a cut below $k_{\min}=4{\times}10^{-3}\,h/\text{Mpc}$. We found adding in smaller-$k$ to the fit did not drastically improve the constraints on $k_0$ and the higher cosmic variance on these large scales increased the chances of worsening the fit. This was especially true for the smaller volume UHF-band results. This also has the added benefit of lessening the potential affect from any primordial non-Gaussianity \citep{Cunnington:2020wdu} (for which we assume none), or foreground systematics, both of which would have their largest influence on the small-$k$ modes we are choosing to cut. 

\begin{figure}
	\centering
  	\includegraphics[width=\columnwidth]{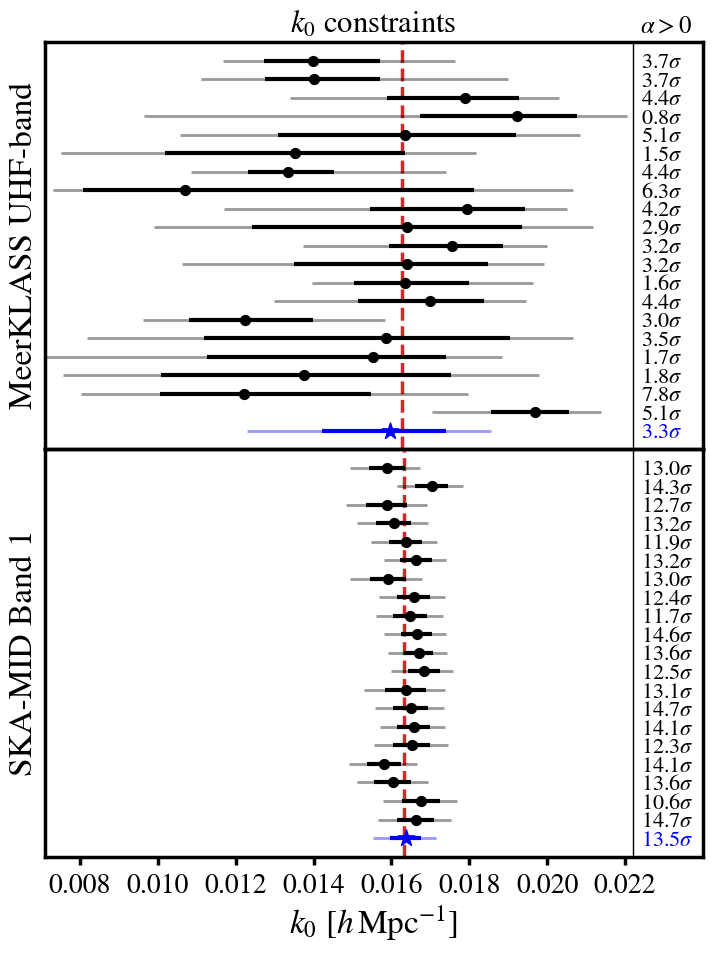}
    \caption{Constraints on the turnover scale $k_0$ from each realisation of simulated data for both the MeerKLASS UHF-band (\textit{top}) and SKA-MID (\textit{bottom}). The simulated data is inclusive of foreground contamination with the transfer function corrections. Constraints are obtained from MCMC results using the logarithmic polynomial fitting technique. The \textit{blue-stars} show the constraint from the averaged power spectrum data from all 20 realisations. The error bars represent 68\% confidence (\textit{dark-thick} lines) and 95\% confidence (\textit{lighter-thin} lines). Final constraints are explicitly outlined in \autoref{tab:k0constraints}. The \textit{right-hand} column of $\sigma$ values shows the turnover ($\alpha{>}0$) detection confidence using the parabola fitting technique for each realisation.}
\label{fig:k0constraints}
\end{figure}

We also examine the performance of the turnover fitting routines on the individual simulated data sets. In \autoref{fig:k0constraints} we show the $k_0$ constraints for MeerKAT and SKAO using the logarithmic polynomial methods. The fiducial $k_0$ is shown by the vertical red dashed line and the data points represent the $k_0$ constraint for each of the 20 realisations with their 1 and $2\sigma$ uncertainties. This visually displays how higher the scatter will be for a MeerKAT survey, yet it is still encouraging to see only two failed (${>}2\sigma$) results. The SKAO data provides much more consistent results with more Gaussian symmetric uncertainties. The blue-starred data point represents the average of all 20 realisations and are thus equivalent to the $k_0$ results already presented in \autoref{fig:SimulationPks}.

To analyse the turnover detection confidence for individual simulations, the right side of the plot shows each realisations turnover detection significance using the parabola fitting technique (\autoref{eq:ParabolaFit}). As previously discussed, despite returning more biased $k_0$ results compared to the logarithmic polynomial technique, it is still a useful method to use for the purpose of quantitatively evaluating the strength of a turnover detection. This is achieved by analysing the value and precision of the parameter $\alpha$ relative to a null model with no turnover ($\alpha{\leq}0$). This shows how the results from simulated data on average agree reasonably with the earlier turnover detection significance calculated on the synthetic model data in \autoref{fig:MKalphk0posterior}. There also appears to be some correlation between the strength of turnover detection and how likely the $k_0$ constraint is to be a biased estimate. Generally,
where there is a low detection significance the estimate for $k_0$ appears more likely to be inaccurate. This makes intuitive sense and can be a useful tool for evaluating the confidence to place in results from a single realisation of real data.

The bottom panel of \autoref{fig:k0constraints} shows the more impressive SKAO results. Here much tighter constraints on $k_0$ are possible and there are far fewer examples of biased and failed estimates, with the scatter of the results generally consistent with the estimated uncertainties. The turnover detection significance quoted in the right-hand column are also overwhelmingly certain of a turnover being present in all realisations. This is unsurprising given the huge scales SKAO will be able to probe. Furthermore, we highlight that for the purpose of the turnover detection confidence alone we did not make a $k_{\min}$ cut on the data and are therefore evaluating the presence of a turnover on the same $k$-range as that presented in the model data in \autoref{fig:SurveyPks}. With this considered, it is therefore expected that an overwhelming detection of a turnover would be made.

\begingroup
\setlength{\tabcolsep}{10pt} 
\renewcommand{\arraystretch}{1.6} 
\begin{table}
	\centering
	\begin{tabular}{lcc}
        \multicolumn{2}{c}{ \textsc{$k_0$ Constraints with \hi\ Intensity Mapping}} \\
        \specialrule{.2em}{.1em}{.1em}
        \textbf{Survey} &
        $k_0\,[h\,\text{Mpc}^{-1}]$& Mean \% error\\
        \hline
		MeerKLASS UHF-band & $0.01596\,_{-0.00173}^{+0.00145}$ & 10.0\%\\
		SKA-MID Band 1 & $0.01638\,_{-0.00041}^{+0.00038}$ & 2.4\% \\
        \specialrule{.2em}{.1em}{.1em}
	\end{tabular}
    \caption{Constraints on turnover scale $k_0$ with 68\% confidence intervals for the simulated \hi\ IM survey data outlined in \autoref{tab:SurveyTable}. These final constraints are from the averaged power spectrum data from 20 simulation realisations and obtained from fitting the cosmology-independent logarithmic polynomial model (\autoref{eq:PolyfitModel}).} 
    \label{tab:k0constraints}
\end{table}

We explicitly state the $k_0$ constraints from MeerKAT's UHF-band and SKAO achieved on the averaged realisations along with $1\sigma$ intervals in \autoref{tab:k0constraints}. We can compare these with the only data constraints available from analysis of the WiggleZ data in \citet{Poole:2012ex}. They achieved a constraint of $0.0160\,_{-0.0035}^{+0.0041}\,h/\text{Mpc}$. Our forecasts suggest MeerKAT may be capable of improving these constraints by nearly a factor of 3, and the SKAO should be able to make an order of magnitude improvement. For a Stage-IV-like spectroscopic survey such as that forecast in \secref{sec:OptComparison} and \autoref{tab:OpticalComparison}, the forecast percentage error is 3\% on a $k_0$ measurement, which as seen from \autoref{tab:k0constraints} will be surpassed by the SKAO.

\subsubsection{A discussion on systematics}\label{sec:Systematics}

We introduced some of the main causes of systematics when establishing the \hi\ power spectrum formalism in \secref{sec:HIIMPower}, but we add some extra discussion here as caveats to some of the results we have presented. Whilst we have included the effects of signal loss from foreground contamination, which is likely to be the most dominant impact on a turnover constraint, there are many other systematic effects which have not been directly included. We included the presence of thermal noise which appeared to have no impact, other than to boost errors. We reasonably assume that other noise-like components such as residual RFI and any non-Gaussian noise should only cause small additive biases, without any extreme scale dependence, thus cause negligible distortion to the turnover. 

One issue we have not considered in great detail is the impact from the telescope beam. We mostly make the assumption that it can be well modelled and thus corrected for. Indeed, we demonstrate this for a single realisation of data with a frequency-dependent cosine-beam with side-lobes in \appref{app:Systematics} and \autoref{fig:BeamModelDemo}. Any imprecision in the beam model should have the most impact on the smallest scales, thus rendering the turnover robust to beam systematics. However, as we also demonstrate in \appref{app:Systematics}, a very incorrect model can still cause biased turnover constraints. The combination of a complex beam with foreground contamination can also cause issues as we have discussed. The main reason for not including this study was due to practical reasons. The cosine-beam model we use is very computationally expensive and would require running on our full suite of simulations, also including it in the mocks for the the transfer function calculation. Then investigating how ``incorrect'' we could allow our beam model to be proved too computationally demanding. Other studies are also addressing in detail the general impact from such issue \citep{Matshawule:2020fjz,Spinelli:2019smg}. See \appref{app:Systematics} for further discussion concerning the beam.

An issue which is likely to be more problematic is the impact from convolution effects from window functions used to account for incomplete survey masking. This was identified in \citet{Poole:2012ex} as the most dominant source of systematic in their attempt to constrain the turnover using the WiggleZ galaxy survey. For \hi\ IM, window functions could potentially be challenging to model due to RFI flagging, foreground cleaning, scanning patterns etc. all of which create a non-uniform signal with a complex window function. \textcolor{black}{As a simple demonstration into their impact, we generated a model \hi\ power spectrum interpolated over a large $20\,{\rm Gpc}^3h^{-3}$ grid, then convolved this with a SKAO survey size top-hat window function. This only caused a small $2.4\%$ shift in the turnover location, but would cause a significant bias for the purposes of precision cosmology. Whilst completely ignoring these masking effects would be unlikely in a robust future analysis, an increased complexity in the true survey window function is inevitable and it is likely that issues in correctly modelling it will arise.} Window functions for \hi\ IM is not well covered in current literature and their affects are thus poorly understood. \textcolor{black}{We therefore do not consider their impact in this work, beyond this simple demonstration, since a more dedicated study into their wider generalised influence is warranted.}

\textcolor{black}{Further limitations beyond systematics could arise from emerging physical phenomena causing degeneracies with a turnover constraint. For example, primordial non-Gaussian fluctuations could be present in our Universe caused by multi-field models of inflation. These create a scale-dependent bias in tracer fields of large-scale cosmic structure, parameterised by $f_{\rm NL}$, which is most sensitive at small-$k$ \citep{Komatsu:2001rj}. A strong presence of this could distort the shape of the power spectrum, and shift the turnover location causing a degeneracy between $f_{\rm NL}$ and $k_0$. However, a simple test on a model \hi\ power spectrum revealed that setting $f_{\rm NL}=5$, only induces a $0.69\%$ shift in the turnover location. So to create beyond $1\%$ shifts in $k_0$ would require an $f_{\rm NL}$ inconsistent with Planck18 data \citep{Akrami:2018mcd}. Relativistic effects \citep{Camera:2014sba,Fonseca:2015laa} are another example of a scale-dependent impact on the power spectrum at large-scales, as are compensated isocurvature perturbations \citep{Hotinli:2019wdp}. For simplicity though, we assume no noticeable contributions are present from these scale-dependent effects at the turnover scales, leaving an investigation into possible degeneracies to future work.}

\section{Cosmology from Turnover Scales}\label{sec:CosmofromTurnover}

In this final section before concluding, we speculate on the possibility of extending this analysis into a more direct probe of cosmological parameters. As discussed in the introduction, the turnover feature is formed based on the horizon size at matter-radiation equality, an epoch in the very early stages of the Universe. This means that changes in a cosmological model can shift the position of the turnover location $k_0$. Thus, assuming our forecasts for the precise measurement on the turnover location $k_0$, competitive constraints on cosmological parameters should be attainable.

Much previous work has investigated how exploiting information contained in the turnover and the full-shape power spectrum on larger scales can break certain degeneracies \citep{Ivanov:2019pdj,Philcox:2020vvt,Baxter:2020qlr,Philcox:2020xbv,DAmico:2020ods}, reveal information about curvature \citep{Vagnozzi:2020rcz} and also sharpen constraints on beyond $\Lambda$CDM models \citep{Chudaykin:2020ghx}. We refer the reader to these investigations for a more comprehensive analysis of this topic in general.



\subsection{Standard rulers from an intensity mapping turnover}\label{sec:StandardRulersIM}

Since the turnover location is determined by the horizon size at matter-radiation equality, it represents a source of geometric information which can in principle be used as a distance measure or "standard ruler". The more typically used standard ruler relates the angular scale of the BAO which can be compared to the theoretical size of the sound horizon at decoupling \citep{Blake:2003rh}. However, as we have discussed, the BAO feature will be a challenging detection for single-dish IM surveys with large beams such as MeerKAT and SKAO. 
We therefore aim to investigate if a detected turnover feature using \hi\ IM, can instead be used as the geometric source for the standard ruler. To introduce the formalism used in this test, we first consider how distance information is extracted from the BAO \citep[see e.g.][]{Aubourg:2014yra,Anderson:2012sa,Bautista:2020ahg}.

A distance to the scale of the observed BAO features can be measured relative to the sound horizon $r_\text{s}$. From the spherically averaged power spectrum the volume averaged distance is utilised, given by \citep{Eisenstein:2005su}
\begin{equation}
    D_{\mathrm{V}}(z)=\left[(1+z)^{2} D_{\mathrm{A}}^{2}(z) \frac{c z}{H_0 E(z)}\right]^{1 / 3}\,,
\end{equation}
where the angular diameter distance is given by
\begin{equation}
    D_\text{A}(z) = \frac{c}{H_0(1+z)} \int_{0}^{z} \frac{\text{d}z^{\prime}}{E(z^{\prime})}\,,
\end{equation}
with $E(z)=\sqrt{\Omega_\text{m}(1+z)^3+\Omega_\Lambda}$, assuming flatness and negligible radiation content in the late Universe. Template functions based on a fiducial cosmology can then be fitted to the scales around the BAO feature with a dilation factor included $P_\text{fid}(k/\alpha_\text{BAO})$. This constrains the scale of the BAO feature relative to a fiducial input. A distance to the redshift of the observed BAO features relative to the sound horizon is then provided through
\begin{equation}
    \alpha_\text{BAO}=\frac{D_\text{V}(z)/r_\text{s}}{\left(D_\text{V}(z)/r_\text{s}\right)^{\mathrm{fid}}}\,.
\end{equation}
If this dilation parameter is a small enough deviation from unity, then it can accurately rescale the fiducial cosmology to what is measured without constant recalculations using Boltzmann codes or perturbation theory models, as required in full-shape measurements \citep{Ivanov:2019pdj}.

When using the turnover as the geometric feature, a similar process can be adopted and we follow the procedure used in \citet{Poole:2012ex} on WiggleZ data. Based on our model-independent fit to the \hi\ IM simulations, and measurement of the turnover scale $k_0$, we can estimate a dilation parameter of the feature scale as
\begin{equation}
    \hat{\alpha}_0 = \frac{k_{0,\text{fid}}}{k_0}\,,
\end{equation}
where we assume a fiducial cosmology to calculate $k_{0,\text{fid}}$. By using a restricted $k$-range around the turnover, and model-independently fitting the peak of the power spectrum, we should be sensitive to information related to matter-radiation equality. This is similar to BAO studies where they aim to obtain information from the BAO-peak, not from the full-shape power spectrum \citep{Kirkby:2013fh}. We are then able to perform a distance measurement to the effective redshift relative to the horizon size at matter-radiation equality $r_\text{eq}$
\begin{equation}
    D_\text{V}(z_\text{eff})/ r_\text{eq} = \alpha_0\left(D_\text{V}(z_\text{eff})/r_\text{eq}\right)^\mathrm{fid}\,.
\end{equation}
We can derive an expression for the horizon size given by \citep{Eisenstein:1997ik,Prada:2011uz}
\begin{equation}
    r_\text{eq}=\frac{4 - 2\sqrt{2}}{\sqrt{2\,\Omega_\text{cb}\,H_0^2\,\zeq}}\,.
\end{equation}
The redshift at matter radiation equality is well constrained by CMB experiments via 
\begin{equation}
    \zeq = 2.5 {\times} 10^4\,\Omega_\text{cb}\,h^2\,\Theta_{2.7}^{-4}\,,
\end{equation}
where $\Theta_{2.7} \equiv T_\text{CMB} /(2.7\,\text{K})$. We use $T_\text{CMB}=2.72548\,\text{K}$ \citep{Fixsen:2009ug}.

Using this process we can then infer a distance to the effective redshift of our simulated data. For the MeerKLASS UHF-band averaged simulations  we obtain a distance constraint of $D_\text{V}(z_\text{eff}{=}0.9)/r_\text{eq} = 26.2_{-2.18}^{+3.19}$ and for the SKA-MID we have $D_\text{V}(z_\text{eff}{=}1.7)/r_\text{eq} = 35.6_{-0.80}^{+0.92}$. The $1\sigma$ confidence achieved on these measurements suggest a matter-radiation equality anchored distance can reach $10.2\%$ for MeerKAT, and $2.4\%$ precision for SKAO.

\subsection{Cosmological parameter estimation with $k_0$}\label{sec:Cosmofromk0}

Using the formalism from the previous section, parameter inference can be run using the information from the turnover. To demonstrate some possible constraints we investigated to what precision the Hubble constant ($H_0$) could be estimated, with all other parameters fixed at their fiducial quantities. Measurements which infer the $H_0$ using the BAO scale as a standard ruler \citep{Aghanim:2018eyx} are currently in tension with \textit{direct} measurements of supernovae and a Cepheid-calibrated local distance ladder \citep{Riess:2019cxk}. This makes it a particular interesting parameter to consider. 

Since the BAO-based $H_0$ constraints have a strong dependence on the sound horizon $r_\text{s}$, the Hubble tension can be equivalently viewed as a sound horizon tension \citep{Bernal:2016gxb,Aylor:2018drw,Knox:2019rjx}. Therefore, the most popular class of solutions for obtaining a $H_0$ measurement more consistent with a local distance ladder approach, involves adjustments to the sound horizon $r_\text{s}$ at decoupling. This has motivated theoretical models such as early dark energy which modify physics at decoupling causing a reduction in the sound horizon \citep[e.g.][]{Poulin:2018cxd}. Recent alternative approaches have sought to probe $H_0$ using LSS in ways independent of the sound horizon, either by imposing priors which break the degeneracy between $r_\text{s}$ and $H_0$ \citep{Pogosian:2020ded}, or by probing larger scales and full-shape power spectrum information, which exploits the $H_0$ dependence on the horizon scale at matter-radiation equality \citep{Ivanov:2019pdj,Philcox:2020vvt,Baxter:2020qlr,Philcox:2020xbv,DAmico:2020ods,Farren:2021grl}. This makes cosmological constraints inferred from the turnover intriguing because the physics which formed the feature occurred before decoupling. Thus distance measures from the turnover scales ($D_\text{V}/r_\text{eq}$) should be very insensitive to the sound horizon, $r_\text{s}$. 

\autoref{fig:H0fromk0contour} demonstrates the posterior distribution on $H_0$ probed using the turnover location $k_0$. To do this we used the MCMC analysis of the turnover feature ran on the simulated data for MeerKLASS and the SKAO. Interestingly, the SKAO data returns a $H_0$ constraint whose $1\sigma$ uncertainty is tight enough that it would also be in significant tension with a current local distance ladder experiment. However, since this represents a constraint which is highly independent of the sound horizon, it means models such as early dark energy, which attempt to adjust the sound horizon, would no longer be appropriate explanations for such tension\footnote{There is already much debate as to whether an early dark energy model is consistent with large scale structure data \citep[see e.g.][]{Hill:2020osr,Ivanov:2020ril,DAmico:2020ods,Smith:2020rxx,Jedamzik:2020zmd}.}.

\begin{figure}
	\centering
  	\includegraphics[width=\columnwidth]{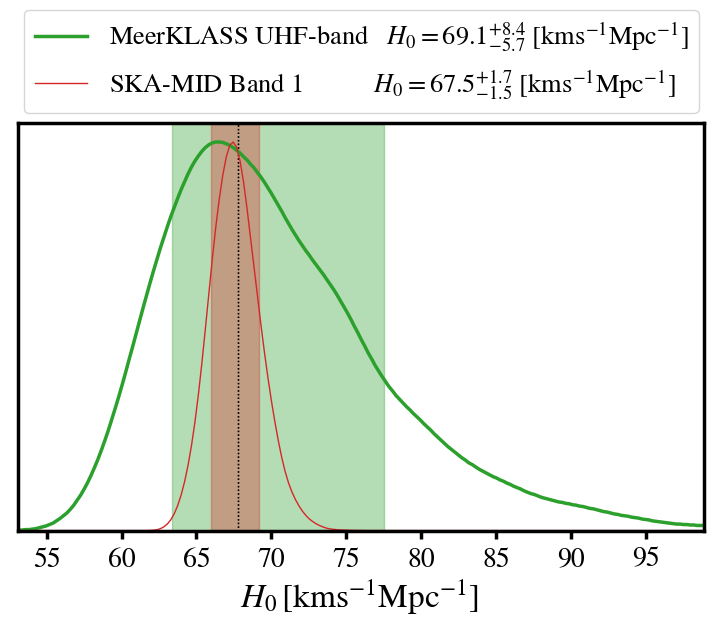}
    \caption{Constraints on $H_0$ from a model-independent logarithmic polynomial fit to turnover scale of the \hi\ power spectrum. By obtaining a dilation parameter, $\alpha_0$, (as discussed in \secref{sec:StandardRulersIM}) an inference on $H_0$ can be obtained from the physics of matter-radiation equality. Black vertical dotted line marks the fiducial $H_0$ quantity and shaded regions indicate the $1\sigma$ confidence regions for both \hi\ IM surveys tested. We have made the simplification that all other parameters are fixed, thus this represents an early demonstration rather than a robust forecast.}
\label{fig:H0fromk0contour}
\end{figure}

The results in \autoref{fig:H0fromk0contour} do not represent a rigorous approach to ensuring a robust constraint. We have made the unrealistic assumption of keeping all other parameters fixed to provide a speculative constraint on $H_0$. A proper analysis would require a more precise construction of the likelihood with carefully selected priors. This may need to account for varying neutrino mass possibilities which will affect the time of the matter-radiation epoch \citep{Kreisch:2019yzn}, and thus the parameter $r_\text{eq}$. Sufficient testing would also be needed to support the claim that the measurements were independent of the sound horizon. Furthermore, a correct treatment for the effect of redshift evolution in such deep redshift bins, or a strategy using multiple thinner redshift bins would need to be implemented. Developing an optimal methodology which addresses all these issues is well beyond the intended aims of this work and we therefore leave this for future investigation.

\section{Conclusions}\label{sec:Conclusion}

As \hi\ IM data starts to arrive, the primary aims will be making a detection in auto-correlation and searching for ways in which it can make unique contributions to cosmological constraints. Probing the turnover feature in the ultra large-scales of the power spectrum can assist in these goals. 

Pathfinder data will have significant contributions from systematics which are not currently well understood. These often cause additive biases to the measured \hi\ power spectrum, and whilst techniques exist for mitigating these, it will be hard to discern what portion of the power spectrum amplitude is from systematic residual or \hi\ signal. This makes any claim of cosmological detection in auto-correlation, based on amplitude alone, particularly challenging. Features in the \hi\ power spectrum such as the turnover provide \textcolor{black}{conclusive evidence} of cosmological signal, and if observed in the correct location, it will strengthen any claim of cosmological detection. Furthermore, since the turnover is related to the horizon size at matter-radiation equality, it can potentially serve as a probe of the  primordial Universe, if a sufficiently precise measurement of its location is achieved. 

In this work we have shown that MeerKAT, a precursor to the SKAO, will be able to detect the turnover feature with a wide 4,000$\,\text{deg}^2$ IM survey. Furthermore, if using MeerKAT's UHF-band, competitive constraints on the turnover location, $k_0$, should be possible. The precision of this constraint is vastly improved with a more advanced survey like a wide 20,000$\,\text{deg}^2$ survey with the full SKAO, which then could allow for competitive cosmological parameter inference.

We summarise our main conclusions below:

\begin{itemize}[leftmargin=*]

\item Using a model-independent parabola fit (\autoref{eq:ParabolaFit}) to \hi\ power spectrum data representative of a MeerKLASS survey, we forecast a detection of the turnover feature with $3.1\sigma$ confidence using MeerKAT's UHF-band. Due to the lower volumes observed, this falls to just under $1\sigma$ for an identical survey in L-band. The details of our survey assumptions were outlined in \autoref{tab:SurveyTable}. This could assist in the confirmation of a successful \hi\ power spectrum which is yet to be achieved with IM in auto-correlation. The detection in UHF-band should surpass all current Stage-III galaxy surveys in optical and near-infrared wavelengths. In the future SKAO will deliver an improved $13.1\sigma$ detection, comfortably double that of a Stage-IV-like spectroscopic survey (see \autoref{tab:OpticalComparison}).
\\
\item Assuming a turnover in power is present, a model-independent logarithmic polynomial (\autoref{eq:PolyfitModel}) can be fitted to the \hi\ power spectrum, from which an accurate estimate of the turnover location $k_0$ can be achieved \autoref{fig:k0histograms}. Our simulation tests provided a constraint of  $k_0=0.01596\,_{-0.00173}^{+0.00145}\,h/\text{Mpc}$ (\autoref{tab:k0constraints}) for a MeerKLASS UHF-band survey. This improves to $k_0=0.01638\,_{-0.00041}^{+0.00038}\,h/\text{Mpc}$ if using the full SKAO. These can both surpass the precision achieved with WiggleZ galaxy data ($0.0160\,_{-0.0035}^{+0.0041}\,h/\text{Mpc}$), and the SKAO should also surpass constraints from a Stage-IV-like spectroscopic survey.
\\
\item We showed how turnover detection and location constraints are robust to signal loss from foreground contamination, the most likely systematic to affect a turnover measurement. We corrected for the distorting effects to the \hi\ power spectrum from foreground cleaning by using transfer functions to estimate the signal loss on the data. Crucially, the mocks used to construct the transfer function appear resistant to an incorrect assumption of the underlying cosmology. We demonstrated this by using a legacy WMAP5 cosmology to construct the mocks for the transfer function, which differs from the Planck18 cosmology used in the simulation of all other data.  
\\
\item We presented arguments for how the turnover's location in $k$-space, means it should, in principle, avoid the strongest influence from most other systematics. Any issues modelling the beam should mostly affect small-scales, and additive biases from noise and residual RFI will naturally effect the turnover the least, where the power amplitude peaks.
\\
\item Since the turnover scale is closely related to the horizon size at matter-radiation equality, it can be used as a standard ruler for calibrating the cosmology-dependent distance-redshift relation. We forecast the precision on a distance to the effective redshift of a MeerKLASS UHF-band survey to be ${\sim}$10\%, which improves to nearly $2\%$ for the SKAO.
\\
\item Combining the distance measurements with information on matter-radiation equality from the CMB, constraints on cosmological parameters can then be inferred. We outlined a simple demonstration of this by placing constraints on the Hubble constant $H_0$ (\autoref{fig:H0fromk0contour}). This would have particularly relevant interest since a favoured solution to resolving the $H_0$ tension is revising early-Universe physics to adapt the sound-horizon scale. Probing $H_0$ using the equality scale should be insensitive to the sound horizon and thus potentially presents a technique to break the $H_0$-sound horizon degeneracy. \newline
\end{itemize}

\noindent In future work we aim to investigate the impact from including further systematics on a turnover measurement with \hi\ IM (as discussed in \secref{sec:Systematics}). We also hope to extend the investigation on the potential for cosmological parameter inference using the turnover with a more sophisticated and robust methodology. Eventually extending the analysis onto a large sky light-cone will be necessary where plane-parallel approximations we have made in this work, will no longer be valid. This will also require a treatment for wide-angle effects. One way to circumvent these issues associated with curved skies is to move from the Fourier power spectrum to a 2D tomographic angular power spectrum approach \citep[e.g.][]{Liu:2016xzv,Camera:2018jys}. Probing a turnover using a harmonic-space power spectrum would thus be an interesting exploration. Lastly, given the similarly optimistic turnover constraints from future photometry surveys (\autoref{tab:OpticalComparison}), an investigation which considers the impact from redshift uncertainty in these experiments would provide a more complete forecast. Incorporating this into a cross-correlation study with \hi\ IM may provide the most optimistic possibility where differing systematics related to both surveys can be mitigated \citep{Alonso:2017dgh,Cunnington:2018zxg,Witzemann:2018cdx,Cunnington:2019lvb,Guandalin:2021sxw}. \textcolor{black}{A further related benefit from cross-correlations would be a reduction in cosmic variance due to the multi-tracer approach \citep{Seljak:2008xr,McDonald:2008sh,Zhao:2021ahg}.}

\section*{Acknowledgements}

SC would like to thank Chris Blake, Jos\'{e} Fonseca, Alkistis Pourtsidou and Mario Santos for useful discussions in the development of this project and Stefano Camera for his review of the manuscript. Furthermore, SC appreciates helpful comments and questions from Julian Bautista, Phil Bull, Isabella Carucci, Eva-Maria Mueller and Daniel Pryer. Lastly, the author is grateful to the referee, whose thorough review improved the quality of the paper. SC is supported by a UK Research and Innovation Future Leaders Fellowship grant [MR/S016066/1] (PI: Alkistis Pourtsidou). This research utilised Queen Mary's Apocrita HPC facility, supported by QMUL Research-IT \url{http://doi.org/10.5281/zenodo.438045}. The use of open source software \citep{scipy:2001,Hunter:2007,mckinney-proc-scipy-2010,numpy:2011} has been used in this project. 

\section*{Data Availability}

 The data underlying this article will be shared on reasonable request to the corresponding author.



\bibliographystyle{mnras}
\bibliography{Bib} 



\appendix






\section{Simulating Systematics}

\subsection{Foreground simulations, cleaning and signal loss reconstruction with transfer functions}\label{app:FGsims}

To simulate the foregrounds, we used the Global Sky Model from the publically available \texttt{PyGSM}\footnote{\href{https://github.com/telegraphic/PyGSM}{github.com/telegraphic/PyGSM}} \citep{PyGSM1,PyGSM2}, which produces full-sky maps covering the emission from $0.01$ to $100\,\mathrm{GHz}$ extrapolated from real data sets. It constructs 
\texttt{HEALPix}\footnote{\href{http://healpix.sourceforge.net}{healpix.sourceforge.net}} maps \citep{Healpy1,Healpy2}, which we then cut the appropriate size from and add to the Cartesian gridded data of the lognormal \hi\ simulations.

As discussed in the main text, we used a PCA process to attempt to clean this contamination. This begins with the covariance matrix of the ``observed'' foreground contaminated data $\mathbf{X}_\text{obs}$ computed by: $\mathbf{C} = \mathbf{X}_\text{obs}^\text{T}\mathbf{X}_\text{obs}/(N_\theta - 1)$. The eigen-decompositon of the covariance matrix, given by $\mathbf{C}\mathbf{V}=\mathbf{V}\mathbf{\Lambda}$, supplies the eigenvectors $\mathbf{V}$ from which the most dominant $N_\text{fg}$ vectors are selected to form the mixing matrix $\mathbf{A}$. The estimated foreground contamination to remove from the data is then calculated with $\mathbf{\widehat{X}}_\text{FG} = \mathbf{A} \mathbf{A}^\text{T} \mathbf{X}_\text{obs}$. We refer to \citet{Cunnington:2020njn} for a more detailed description and dedicated tests of this process. We also note that PCA represents arguably the most basic form of blind foreground cleaning. Many other more sophisticated methods haven been experimented with \citep{Wolz:2013wna,Shaw:2014khi,Carucci:2020enz,Makinen:2020gvh,Fonseca:2020lmi,Soares:2021ohm,Irfan:2021bci}. Comparisons between many of these are presented in \citet{Spinelli:2021emp}. However, PCA is still the most tested on real data and we thus stick to this technique. 

For cleaning the foreground contaminated simulations, we found removing $N_\text{fg}=4$ principal components to be sufficient. However, this unavoidably results in some signal loss to the cosmological-\hi, typically at large scales. As discussed in the \secref{sec:FGtests}, we employ a foreground transfer function to correct for this signal loss, otherwise the shape of the \hi\ power spectrum becomes distorted resulting in a highly biased measurement of the turnover scale $k_0$.

We construct the transfer function by adding mock data $\mathbf{M}$ to the main simulations inclusive of the foreground contamination which we treat as the ``true observed data'' $\mathbf{X}_\text{obs}$. This can then be cleaned to provide $\mathbf{M}_\text{cleaned}$, an estimate for the effects of removing the foregrounds on the mock map:
\begin{equation}\label{eq:MockforTF}
	\mathbf{M}_\text{cleaned} = [\mathbf{M} + \mathbf{X}_\text{obs}]_\text{PCA} - [\mathbf{X}_\text{obs}]_\text{PCA}\, .
\end{equation}
where the $[\ ]_\text{PCA}$ represents an operator which performs a PCA clean with the same number of components removed i.e. $N_\text{fg}=4$. The cleaned data $[\mathbf{X}_\text{obs}]_\text{PCA}$ is also subtracted in \autoref{eq:MockforTF} which is necessary to reduce the extra unnecessary variance caused by the presence of data-\hi\ in the mock signal. The transfer function is then given by:
\begin{equation}\label{eq:TransferFunc}
	T(k) = \left\langle  \frac{\mathcal{P}(\mathbf{M}_\text{cleaned}\, ,\, \mathbf{M})}{\mathcal{P}(\mathbf{M}\, ,\, \mathbf{M})} \right\rangle \, ,
\end{equation}
where $\mathcal{P}()$ denotes an operator which measures the power spectrum with the same binning assumed for the real data. The angled brackets denote an averaging over a large number of mocks. The power spectrum is then corrected for by dividing through by this transfer function. This has been used on real data \citep{Masui:2012zc,Anderson:2017ert,Wolz:2021ofa} and also analysed in simulations \citep{Cunnington:2020njn}.

Since mock data needs to be injected into the observed data to construct the transfer function, it would be very problematic if the cosmology assumed for this mock data had a large impact on the signal reconstruction. This is the motivation behind modelling the signal loss with phenomenological functions instead of correcting for it with transfer functions, since then nuisance parameters in such models can be marginalised over in a more statistically robust way. However, this can lead to degeneracies between signal loss and cosmological information on large scales such as constraints on primordial non-Gaussianity \citep{Cunnington:2020wdu}.

\begin{figure}
	\centering
  	\includegraphics[width=\columnwidth]{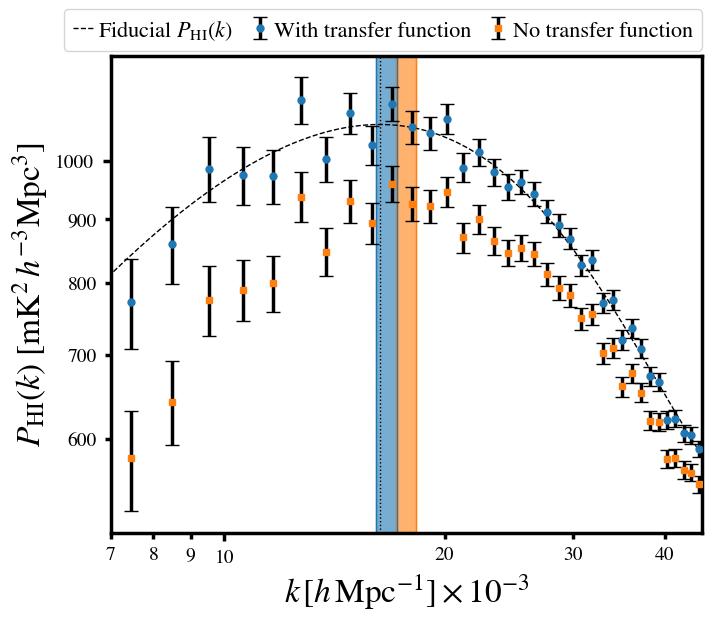}
    \caption{Constraints on the turnover scale for a single-realisation of simulated SKA-MID data with foreground contamination, cleaned using PCA. The \textit{blue-circle} data points show how we have successfully corrected for signal loss with a transfer function, which even assumed an incorrect cosmology. \textit{Orange-square} data points show the effects from signal loss where no transfer function is used. The $1\sigma$ confidence regions on the estimated turnover scale are displayed by the vertical shaded colour regions, which show how using the transfer function allows an unbiased recovery of the fiducial turnover scale (\textit{black-dotted} vertical line.)}
\label{fig:TransferFuncDemo}
\end{figure}

To test whether the turnover measurement is robust to the cosmology used for the transfer function mocks, we took the approach of assuming an incorrect cosmology in the mock generation. Despite using a Planck18 cosmology $\{\Omega_\mathrm{m}, \Omega_\mathrm{b}, h, n_\mathrm{s}\} = \{0.315,0.0489,0.674,0.965\}$ \citep{Aghanim:2018eyx} to generate the simulated observed \hi\ data, for the construction of the mocks in the transfer function we used a different cosmology based on legacy WMAP5 results $\{\Omega_\mathrm{m}, \Omega_\mathrm{b}, h, n_\mathrm{s}\} = \{0.25,0.045,0.73,0.99\}$ \citep{WMAP:2008lyn}. \textcolor{black}{The WMAP5 cosmology has a turnover located at $k_0=0.0153\,h/\text{Mpc}$ relative to the Planck18 $k_0=0.0163\,h/\text{Mpc}$, a ${\sim}6\%$ difference.} This should emulate some ignorance in a real survey and the potential consequences of such ignorance. Fortunately, for the purposes of turnover constraints, we found this did not lead to any issues and were able to make a successful reconstruction of the \hi\ power spectrum, despite this incorrect assumption of the cosmology, as shown by \autoref{fig:TransferFuncDemo}. The blue-circle data points represent a single realisation of SKAO foreground contaminated data, which we PCA clean, then construct and apply the transfer function under the incorrect assumption of WMAP5 cosmology. This recovers the shape of the \hi\ power spectrum (black dashed line) and allows an \textcolor{black}{un}biased recovery of the fiducial turnover scale (black-dotted vertical line). The consequences from not applying a transfer function are shown by the orange-square data points, which demonstrate the distortion on the shape of the power spectrum caused by the signal loss, which leads to biased turnover constraint.

This is evidence to suggest that the transfer functions used are somewhat resilient to the cosmology assumed in their construction. However, a much more dedicated investigation would be needed to support this claim for a range of cosmological applications and a range of differing cosmologies.

\subsection{Incorrect modelling of the beam and other systematics}\label{app:Systematics}

The telescope beam pattern should be something that is reasonably well understood \citep{2021MNRAS.502.2970A} and since its effects are mainly concentrated on small scales, it should not pose much of a problem for constraining the turnover. However, we still found that a very poor model of the beam has the potential to cause issues. To demonstrate this point we looked at including the effects from the telescope beam in the simulated data, then use two cases where the model of the beam was the correct frequency-dependent cosine beam model with side-lobes\footnote{For the cosine beam we follow the same model outlined in \citet{Matshawule:2020fjz} and \citet{Cunnington:2021czb}.} and another incorrect model where we use a simple Gaussian beam with no side-lobes. We show these results in \autoref{fig:BeamModelDemo}.

\begin{figure}
	\centering
  	\includegraphics[width=\columnwidth]{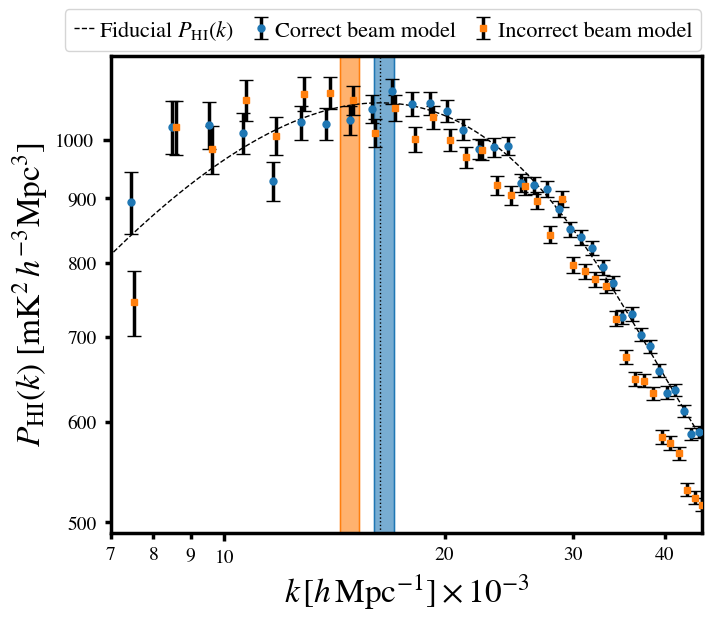}
    \caption{Constraints on the turnover scale for a single-realisation of simulated SKA-MID data including a frequency-dependent cosine beam with side-lobes. The \textit{blue-circle} data points show how we have successfully corrected for the beam when using the correct model. \textit{Orange-square} data points show the effects from where a very incorrect beam model is used which assumed a Gaussian frequency-independent beam without side-lobes. The $1\sigma$ confidence regions on the estimated turnover scale are displayed by the vertical shaded colour regions, which show how an incorrect beam model can cause a biased recovery of the fiducial turnover scale (\textit{black-dotted} vertical line.)}
\label{fig:BeamModelDemo}
\end{figure}

The blue-circle data points are where we have used the correct beam model to reverse the damping effects caused by the beam and recover the correct shape of the fiducial \hi\ power spectrum. This then allows for an unbiased recovery of the turnover scale shown by the blue shaded vertical bar which represents the $1\sigma$ confidence region, which is in agreement with the fiducial $k_0$ (black vertical dotted line). The orange-square data points show the results from assuming an incorrect beam model. Despite the beam effects mainly being concentrated on small scales, this can still cause enough distortion to the \hi\ power spectrum, leading to a biased constraint on the turnover. It is interesting that this seems to cause even more bias than the signal loss from foreground contamination. However, this is likely because we are assuming an SKAO survey here, which at its lowest frequency will have a very large ($3.27\,\text{deg}$) beam. We are also deliberately choosing a very incorrect model to demonstrate the consequences. In reality the beam, should be something that is well understood and can thus be well modelled. However, understanding the relation between the foregrounds and the beam is something that could become very complex. We leave an investigation into the potential impact from a coupling between complex beams and foregrounds on a turnover constraint to future work \citep[see][for a detailed discussion]{Matshawule:2020fjz,Spinelli:2021emp}.

Testing resilience against other systematic effects is a similarly difficult task when using simulated data. This is because the model of the systematics used in the simulated data is then fully known and can therefore be included in the fitting process to correct for the effects with near perfect precision. This is somewhat demonstrated in \autoref{fig:BeamModelDemo} where we display the modelling of two extremes from full knowledge of the systematic to a very naive assumption. A study into how precise the models of systematics need to be for general \hi\ IM precision cosmology, deserves a dedicated study and is far beyond the scope of this work. Many systematics also require more detailed models to even include in simulations such as RFI and masking effects from incomplete survey coverage \citep[see][for a study into RFI from global navigation satellites]{Harper:2018ncl}. Both of these examples could be relevant for successful evaluation of the turnover.


\bsp	
\label{lastpage}
\end{document}